\documentclass{aa}

\usepackage{amssymb}
\usepackage{amsmath}
\usepackage{graphicx}
\usepackage{url,natbib,twoopt}
\usepackage[varg]{txfonts}
\usepackage{subfigure}
\usepackage{todo}
\usepackage{hyperref}

\hypersetup{
    colorlinks=true,
    linkcolor=blue,
    filecolor=magenta,
    urlcolor=blue,
    citecolor=blue,
}

\bibpunct{(}{)}{;}{a}{}{,}

\newcommand{\nustar}{\textit{NuSTAR}\xspace}
\newcommand{\fpma}{FPMA\xspace}
\newcommand{\fpmb}{FPMB\xspace}
\newcommand{\gx}{GX\,304$-$1\xspace}
\newcommand{\xper}{X\,Persei\xspace}
\newcommand{\aplus}{A\,0535+262\xspace}
\newcommand{\gro}{GRO\,J1008$-$57\xspace}
\newcommand{\polcap}{\texttt{polcap}\xspace}

\newcommand{\asec}{\ensuremath{''}\xspace}
\newcommand{\kev}{\ensuremath{\,\text{keV}}\xspace}
\newcommand{\cyc}{\ensuremath{\text{cyc}}}
\newcommand{\gsec}{\,\mathrm{g}\,\mathrm{s}^{-1}}
\newcommand{\gcm}{\,\mathrm{g}\,\mathrm{cm}^{-2}}
\newcommand{\ns}{\mathrm{NS}}
\newcommand{\msun}{\,M_{\odot}}
\newcommand{\diff}{\ensuremath{\mathrm{d}}}

\begin{document}

\title{X-ray emission from magnetized neutron star atmospheres at low mass accretion rates}

\subtitle{I. Phase-averaged spectrum}

\author{
        E.~Sokolova-Lapa\inst{\ref{inst1}, \ref{inst2}
        \thanks{\email{ekaterina.sokolova-lapa@fau.de}}}
        \and M.~Gornostaev\inst{\ref{inst2}}
        \and J.~Wilms\inst{\ref{inst1}}
        \and R.~Ballhausen\inst{\ref{inst1}}
        \and S.~Falkner\inst{\ref{inst1}}
        \and K.~Postnov\inst{\ref{inst2}, \ref{inst3}}
        \and P.~Thalhammer\inst{\ref{inst1}}
        \and F.~F\"urst\inst{\ref{inst4}}
        \and J.~A.~Garc\'ia\inst{\ref{inst1}, \ref{inst5}}
        \and N.~Shakura\inst{\ref{inst2}, \ref{inst3}}
        \and P.~A.~Becker\inst{\ref{inst8}}
        \and M.~T.~Wolff\inst{\ref{inst9}}
        \and K.~Pottschmidt\inst{\ref{inst6}, \ref{inst7}}
        \and L.~H\"arer\inst{\ref{inst1}}
        \and C.~Malacaria\inst{\ref{inst10}, \ref{inst11}}
}

\institute{Dr.~Karl Remeis-Observatory \& ECAP, University of Erlangen-Nuremberg,
Sternwartstr.~7, 96049 Bamberg, Germany\label{inst1}
\and
Sternberg Astronomical Institute, M.~V.~Lomonosov Moscow State University,
Universitetskij pr., 13, Moscow 119992, Russia\label{inst2}
\and
Kazan Federal University, 420008 Kazan, Russia\label{inst3}
\and
European Space Astronomy Center (ESA/ESAC), Science Operations Department,
Villanueva de la Cañada, 28691, Madrid, Spain\label{inst4}
\and
Cahill Center for Astronomy and Astrophysics, California Institute of Technology,
Pasadena, CA 91125, USA\label{inst5}
\and
CRESST, Department of Physics, and Center for Space Science and Technology,
UMBC, Baltimore, MD 21250, USA\label{inst6}
\and
NASA Goddard Space Flight Center, Astrophysics Science Division, Greenbelt, MD
20771, USA\label{inst7}
\and
Department of Physics \& Astronomy, George Mason University,
Fairfax, VA 22030-4444, USA\label{inst8}
\and
Space Science Division, Naval Research Laboratory,
Washington, DC 20375-5352, USA\label{inst9}
\and
NASA Marshall Space Flight Center, NSSTC, 320 Sparkman Drive, Huntsville, AL 35805, USA\label{inst10}
\and
Universities Space Research Association, Science and Technology Institute, 320 Sparkman Drive, Huntsville, AL 35805, USA\label{inst11}
}

\abstract{Recent observations of X-ray pulsars at low luminosities
  allow, for the first time, to compare theoretical models for the
  emission from highly magnetized neutron star atmospheres at low
  mass accretion rates
  ($\dot{M}\lesssim10^{15}\,\mathrm{g}\,\mathrm{s}^{-1}$) with the
  broadband X-ray data. The purpose of this paper is to
  investigate the spectral formation in the neutron star atmosphere
  at low $\dot{M}$ and to conduct a parameter study of physical
  properties of the emitting region.
  We obtain the structure of the static atmosphere, assuming that
  Coulomb collisions are the dominant deceleration process. The upper
  part of the atmosphere is strongly heated by the braking plasma,
  reaching temperatures of 30--40\,keV, while its denser isothermal
  interior is much cooler (${\sim}$2\kev).
  We numerically solve the polarized radiative transfer in the
  atmosphere with magnetic Compton scattering, free-free processes,
  and non-thermal cyclotron emission due to possible collisional
  excitations of electrons.
  The strongly polarized emitted spectrum has a double-hump shape
  that is observed in low-luminosity X-ray pulsars. A low-energy
  ``thermal'' component is dominated by extraordinary photons that
  can leave the atmosphere from deeper layers due to their long
  mean free path at soft energies. We find that a high-energy
  component is formed due to resonant Comptonization in the
  heated non-isothermal part of the atmosphere even in the absence
  of collisional excitations. The latter, however, affect the ratio
  of the two components.
  A strong cyclotron line originates from the optically thin,
  uppermost zone.
  A fit of the model to \textit{NuSTAR} and \textit{Swift}/XRT
  observations of \gx provides an accurate description of the
  data with reasonable parameters. The model can thus reproduce
  the characteristic double-hump spectrum observed in low-luminosity
  X-ray pulsars and provides insights into spectral
  formation.
  }


\date{Received DATA / Accepted DATA }

\maketitle

\section{Introduction}\label{sec:intro}
    
Accretion-powered X-ray pulsars have been known for decades as binary
systems which consist of a rotating highly magnetized neutron star and
a donor star companion \citep[e.g.,][]{zeldovich1969,basko1975}. The
strong magnetic fields of the neutron stars in High-Mass X-ray
binaries (HMXBs) significantly affect the accretion of matter from the
donor, which is typically an early-type (O/B) star. Close to the
compact object, in the magnetosphere of the neutron star, the dynamics
of the accretion flow are dominated by the magnetic field of the
neutron star. The accreting material couples to the magnetic field
lines and is channeled down onto the magnetic poles of the neutron
star. The matter is decelerated from its free-fall velocity of
${\sim}0.7c$ in an accretion column or at the surface of the neutron
star. The properties of the column depend on the mass accretion rate
\citep[e.g.,][]{burnard1991, becker2012, postnov2015, mushtukov2015}.
It is well understood that at high accretion rates of
$\dot{M} \gtrsim 10^{17}\gsec$ radiation pressure plays a major role
in decelerating the matter \citep{davidson1973, basko1976,
  wang1981} in the accretion column. The accretion flow loses its
energy gradually, by passing through the extensive radiative shock.
See \citet{becker2007}, \citet{wolff2016}, \citet{farinelli2016},
and \citet{gornostaev2021} for accretion column models in this
high mass accretion rate regime. At lower accretion rates,
$\dot{M}\sim10^{15}\mbox{--}10^{17}\gsec$, it is possible that matter
is decelerated by passing through a collisionless shock
\citep{langer1982, bykov2004, vybornov2017}. In this case, the gas
shock is discontinuous with post-shock velocity ${\sim}v_\mathrm{ff}$.
The aftershock flow is decelerated to rest by Coulomb collisions.
Soft thermal X-ray photons emitted from the thermalized region near
the surface of the neutron star, as well as photons produced by
free-free and cyclotron emission, are then reprocessed in the column
by Compton scattering to produce the harder X-rays emerging from the
column. At even lower mass accretion rates, the accretion flow
approaches the neutron star surface at free-fall velocity and then
rapidly decelerates in the nearly static atmosphere by Coulomb
collisions \citep[][ZS69 hereafter]{zeldovich1969}.

The extreme magnetic field of the neutron star significantly affects
both the hydrodynamical characteristics of the accretion flow and the
photon-electron interactions inside the plasma. In the presence of
strong magnetic fields, photon scattering, absorption, and emission
become resonant processes due to the quantization of the electron
motion perpendicular to the magnetic field on Landau levels. As a
result, the cross sections for magnetic Compton scattering and
magnetic free-free absorption are strongly energy and polarization
dependent and highly anisotropic \citep{herold1979, nagel1980,
  daugherty1986, bussard1986, schwarm2017a}. The resonant nature of
the scattering process is also responsible for the formation of
Cyclotron Resonance Scattering Features (CRSFs), commonly referred to
as cyclotron lines, which are observed in the spectra of about 36
X-ray pulsars and usually appear as absorption line-like spectral
features \citep{staubert2019}. Besides the fundamental feature located
at the energy corresponding to the electron cyclotron frequency,
$\hbar\omega_\cyc$, within the line forming region, higher harmonics
can also sometimes be observed. The shapes of these lines can differ
from a Voigt profile due to the non-constant magnetic field over the
line forming region and multiple photon Compton scattering, which
creates spawned photons from higher harmonics with energies close to
the fundamental one \citep{schwarm2017a,schoenherr2007a}. A second
effect of the strong magnetic field is the polarization of the
propagating radiation by magnetized plasma and vacuum birefringence.
Due to expected strong Faraday depolarization, the photon field can be
described in terms of two normal modes, ordinary and extraordinary
\citep{gnedin1978}, which exhibit very different properties in
photon-electron interactions. The cross sections for the extraordinary
mode are strongly energy dependent, whereas for the ordinary mode they
are highly anisotropic.

At the sufficiently low accretion rates considered here
($\dot{M}\lesssim10^{15}\gsec$), the picture of the accretion process
is different from one, corresponding to an accretion column. The
formation of radiation-dominated shocks in the tenuous flow is not
possible due to the short radiation diffusion time \citep{postnov2015}.
Instead, matter reaches the surface in free-fall regime, and Coulomb
interactions between particles of the falling plasma and particles of
the neutron star atmosphere are expected to be the main mechanism of
plasma deceleration \citep{harding1984, miller1987}. As protons carry
most of the energy of accreting flow, the main energy transfer occurs
due to falling proton collisions with ambient atmospheric electrons.
The physical picture of this braking of the plasma through Coulomb
collisions and the associated formation of the spectrum of emerging
radiation has been investigated in a number of papers, starting with
ZS69.

ZS69 solved the energy balance and hydrostatic equilibrium equations,
together with the diffusion equation for the energy density, obtaining
the thermal and density structure of the atmosphere without the magnetic
field and for spherical accretion. They first showed that
the upper, optically thin part of such atmospheres is overheated by
the deceleration of accreted matter, while the deep layers have a
much lower electron temperature, $T_\mathrm{e}$. This result was
later confirmed and extended by,
e.g., \citet{alm1973}, \citet{turolla1994}, \citet{deufel2001}, and
\citet{suleimanov2018}. The proton stopping in a magnetized plasma
differs from the non-magnetic case due to the restriction of the
momentum transfer in the perpendicular direction. The protons usually
occupy very high Landau levels and thus can be treated classically.
However, the quantization of the electron's motion affects the Coulomb
collisions and forces protons to veer from the magnetic field lines
\citep{miller1987}. Collisional deceleration in a highly magnetized
plasma is therefore less efficient \citep{harding1984, miller1987}.
However, Coulomb collisions during plasma braking can also lead to the
excitation of electrons to higher Landau levels. These excitations are
then followed by radiative decay, increasing the number of cyclotron
photons in the atmosphere \citep{bussard1980}. To a high degree, this
process depends on the magnetic field strength and the velocity of the
falling matter. For high magnetic fields, when only 1--2 Landau levels
can be collisionally populated, only a small fraction of the accretion
energy goes into non-thermal Landau excitations \citep{miller1987,
  miller1989}. \citet{nelson1993, nelson1995} considered the effects
of a moderate magnetic field, allowing for excitations to very high
Landau levels with subsequent radiative decay. They suggested that due
to down-scattering in the cold atmosphere the escaping flux of these
spawned photons can be observed as a broad emission line-like feature
below the cyclotron resonance energy.

Motivated by recent \nustar observations of the transient X-ray
pulsars \gx \citep{tsygankov2019a} and \aplus \citep{tsygankov2019b},
which allow the direct study of the broadband X-ray spectrum emitted by
highly magnetized neutron stars at low $\dot{M}$, in this paper we
consider the spectral formation in the atmosphere of a neutron star.
Specifically, the \nustar observations were performed at luminosities
of ${\sim}10^{34}\,\mathrm{erg}\,\mathrm{s}^{-1}$ for \gx and
${\sim}7\times10^{34}\,\mathrm{erg}\,\mathrm{s}^{-1}$ for \aplus. Both
sources showed a drastic spectral transition from the exponential
cutoff power law widely observed in HMXBs at intermediate to high
luminosities to a double-hump shape. The soft energy component peaks
at around 5\kev followed by a dip at ${\sim}$10--20\kev and a subsequent
hump, peaking at around 30--40\kev. The analysis of the spectrum of
\aplus by \citet{tsygankov2019b} also showed a possible cyclotron
line at 47.7\kev. The persistent low-luminosity HMXB \xper and the
transient HMXB \gro, which has recently been observed at
${\sim}10^{35}\,\mathrm{erg}\,\mathrm{s}^{-1}$ \citep{lutovinov2021},
are the other examples of a similar peculiar two-component continuum.
\citet{tsygankov2019a} suggested that the high-energy component can be
explained by the production of the cyclotron photons due to
collisional excitations of the Landau levels and their subsequent
reprocessing in the heated atmosphere.

\citet{mushtukov2020} showed that if one assumes a strong primary
source of cyclotron photons and reprocesses these in the upper, hot
part of the atmosphere, the two humped spectrum can be reproduced.
This ansatz requires that almost all accretion flow energy goes into
collisional excitations of atmospheric electrons and the subsequent
production of the cyclotron photons \citep[Eq.~3 of][]{mushtukov2020}.
The latter is, however, not supported by existing models for
low-$\dot{M}$ accretion in strong magnetic fields
\citep[e.g.,][]{miller1989}. In this paper, we revisit
the spectral formation from the polar caps of magnetized neutron
stars by improving on the modeling of the radiative transfer through
the neutron star atmosphere. We obtain the structure of the neutron
star atmosphere
by adopting the approach of ZS69 for non-magnetic atmospheres
and then calculate the polarized radiative transfer. We show that
these calculated synthetic spectra have the potential for widespread
use as a tool for data analysis of low luminosity observations.

The paper is organized as follows. In Sect.~\ref{sec:model} we
discuss the polar cap model, presenting our formulation of the
problem. We describe the atmospheric structure calculation, the
opacities due to Compton scattering and free-free absorption in
a strong magnetic field, and our approach to compute the radiative
transfer in this setup. In Sect.~\ref{sec:results} we then describe
the results of solving the radiative transfer equation and investigate
the parameter dependencies. In Sect.~\ref{sec:gx304-1} we apply the
model to fit the low luminosity observation of the X-ray pulsar
\gx mentioned above. We summarize our results, discuss the
limitations of our model, and outline possible future developments
in Sect.~\ref{sec:disc}.

\section{Polar cap model}\label{sec:model}
\subsection{Overview}\label{ssec:model}

In this section, we describe the polar cap model (\polcap hereafter)
in greater detail. We consider the accretion of fully ionized
hydrogen plasma with a low mass accretion rate,
$\dot{M}\lesssim10^{15}\gsec$, onto a magnetic pole of a neutron
star with mass $M_\ns=1.4\msun$ and radius $R_\ns=12\,\mathrm{km}$.
We assume that under such conditions, neither a radiative nor a gas
mediated shock raises above the surface, i.e., no accretion column
is formed. The matter thus falls freely along the magnetic field
lines. The braking is mediated mainly by Coulomb
collisions within the thin layer of already accreted plasma that
spreads around the magnetic pole and merges into the neutron star
atmosphere. In the following, we refer to this layer as the
``atmosphere''. The accreted energy is reprocessed in the atmosphere
and results in the X-ray emission from the polar cap of radius
$r_0$. At such low accretion rates, we can ignore the influence of
the rarefied incoming plasma on the emergent radiation. The plasma
braking depends to a high degree on the $\dot{M}$ and the surface
magnetic field strength $B_0$. While we do not calculate the details
of the stopping process, we take into account that a fraction of
the accreted energy, $f_\cyc$, can go into collisional excitations
of ambient atmospheric electrons to the higher Landau levels, while
the rest contributes to the thermal energy of the accretion region.

Based on this physical picture, we build our model for the polar cap
emission characterized by four major parameters: the mass accretion
rate, $\dot{M}$; the polar cap radius, $r_0$; the surface magnetic
field strength $B_0$, which we express in terms of the corresponding
cyclotron energy, $E_\cyc$; and the fraction of accretion energy
going into non-thermal collisional excitation of electrons, $f_\cyc$.
In the following, we give a brief overview of the general concept of
our modeling.

We start our discussion in Sect.~\ref{ssec:atm} by describing the 
computation of the inhomogeneous thermal and density structure of
the atmosphere. There, we mainly follow the hydrostatic approach
of ZS69 and only include a rudimentary model for the radiation field.

In a second step, in Sect.~\ref{ssec:sctabs}, we consider the physics
of photon-electron interactions and corresponding opacities in the
highly magnetized medium of the atmosphere.

In Sect.~\ref{ssec:radtran}, we then describe the transfer of the
radiation through the atmosphere using the obtained structure and
opacities. We solve the radiative transfer equation in two
polarization modes, taking into account the anisotropy of the
radiation field and the partial energy redistribution due to magnetic
Compton scattering. The major source of photons throughout the
atmosphere is free-free emission (bremsstrahlung) where we also
account for the presence of a magnetic field. A further source of
radiation is the cyclotron emission due to radiative decay of
electrons excited by non-thermal collisions.

Finally, in Sect.~\ref{ssec:calc} we describe a complete simulation
that combines all the different aspects to obtain a consistent
model for the atmospheric structure and the spectrum of the
emergent radiation.

\begin{figure}
    \resizebox{\hsize}{!}{\includegraphics{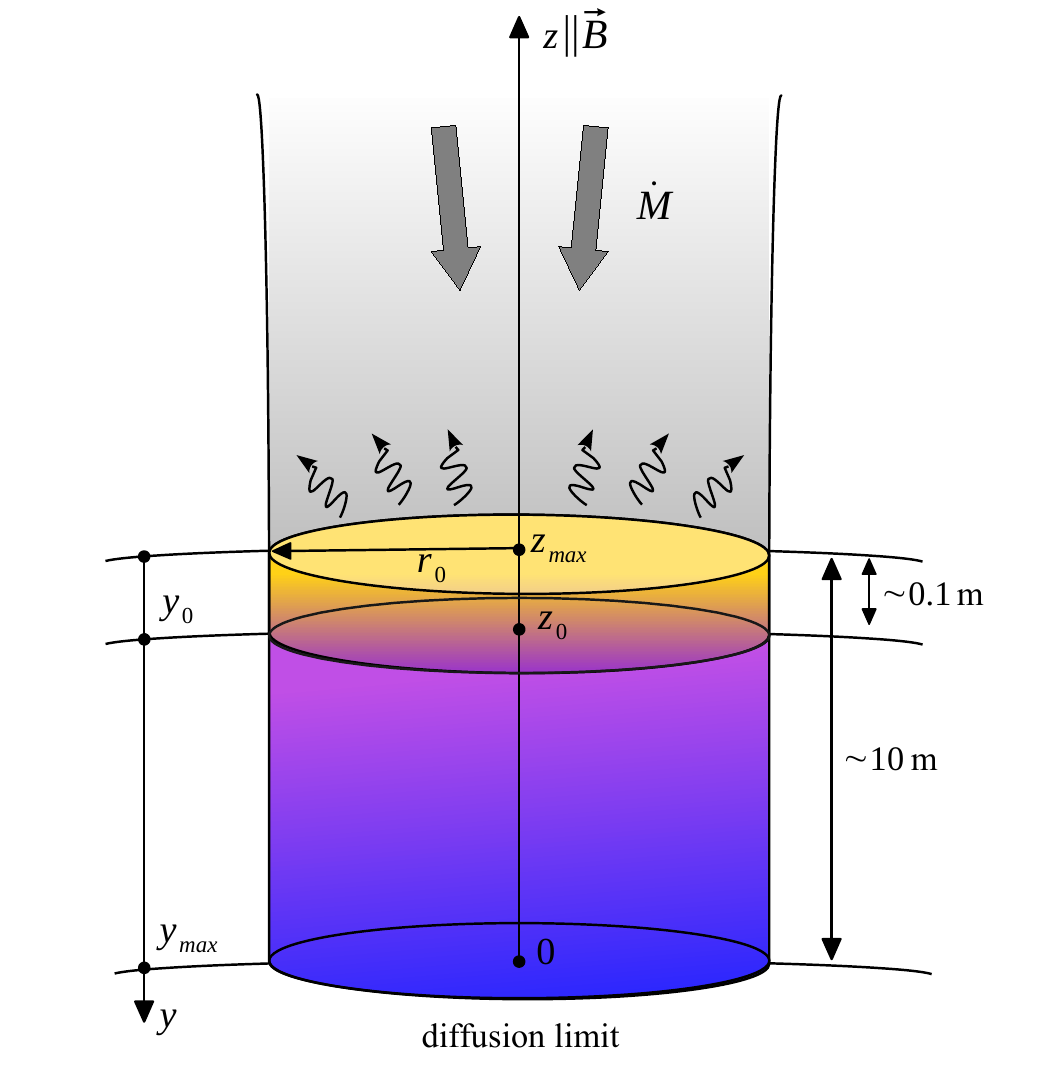}}
    \caption{Schematic depiction of the model describing the
             accretion at low $\dot{M}$ onto the polar cap of a 
             magnetized neutron star. The polar cap is an emitting
             layer of an atmosphere with the total column density
             $y_\mathrm{max}$ (corresponding to the total
             geometrical depth $z_\mathrm{max}\sim10\,\mathrm{m}$)
             and radius $r_0$. The characteristic proton stopping
             length is denoted as $z_0$ (geometrical scale) and
             $y_0$ (column density scale).}
    \label{fig:sketch}
\end{figure}

\subsection{Atmosphere model}\label{ssec:atm}

We assume that the neutron star atmosphere is a plane-parallel
slab. Its outer surface normal is co-directional with the $z$-axis,
which is defined to be parallel to the magnetic field lines near
the surface.

The structure of the atmosphere is mainly determined by the energy
release of the infalling material, i.e., the deceleration of protons
of an accretion flow in the neutron star atmosphere. The different
mechanisms that are responsible for this deceleration have been
studied extensively over the past 50 years \citep[e.g., ZS69;][and
references therein]{basko1975, kirk1982, harding1984, miller1987, nelson1993}.
The important parameter describing the process is the characteristic
depth inside the atmosphere at which the majority of the injected
particles have released their kinetic energy, $z_0$.

The stopping process depends most strongly on the column density
\begin{equation}\label{eq:coldens}
     y(z) = \int_{z}^{z_\mathrm{max}}\rho(z') \diff z',
\end{equation}
where $z_\mathrm{max}$ is the full geometrical depth of the
atmosphere and where $\rho(z)$ is the density of the plasma. We
therefore use this quantity, rather than $z$, to model spatial
variations. We emphasize that even though the total column density
of the atmosphere, $y_\mathrm{max}$, is high (in our typical setup,
$y_\mathrm{max}=10^3\gcm$), the corresponding
$z_\mathrm{max}$ is only ${\sim}10\,\mathrm{m}$ due to high density.
This allow us to consider the magnetic field as constant inside the 
atmosphere. The accreted energy is released down to $y_0 = y(z_0)$.
We refer to $y_0$ as the ``stopping length'' in the following,
adopting the standard terminology from the literature mentioned
above. Figure~\ref{fig:sketch} provides a sketch of the geometry.

Since the atmosphere is in hydrostatic equilibrium, we can compute
its density from the known temperature profile, i.e., from the energy
balance in the atmosphere. We do not perform a detailed computation of
the actual energy loss rate through the atmosphere, $\diff E/\diff y$,
but adopt a modified approach of ZS69 that ensures the continuity of
the heating function throughout the atmosphere. We leave a deeper
discussion of the energy loss rate for Sect.~\ref{dis:atm}.

In the volume above $y_0$, where the protons are decelerated, the
conditions are likely sufficient to permit the collisional
excitation of electrons into higher Landau levels. We parameterize
this complex process by assuming that a fraction of the accreted
energy, $f_\cyc$, goes into non-thermal collisional excitation of
electrons, while the remaining fraction, $1-f_\cyc$, directly heats
the atmospheric plasma.

The main cooling processes in the atmosphere are free-free emission
and Compton cooling. Similar to ZS69 (their Eq.~1.3), we parameterize
the cooling rate for free-free processes as
\begin{equation}\label{eq:lambb}
    \Lambda_\mathrm{B}=5\times 10^{20} \sqrt{T_\mathrm{e}}
    \rho^2\left(1 -\frac{T'}{T_\mathrm{e}}\right)
    \, \mathrm{erg}\,\mathrm{s}^{-1}\,\mathrm{cm}^{-3},
\end{equation}
and the cooling rate due to Compton scattering as
\begin{equation}\label{eq:lambc}
    \Lambda_\mathrm{C}=\frac{4\epsilon c \sigma_\mathrm{T}\rho}
    {m_\mathrm{p}}\frac{kT_\mathrm{e}}{m_\mathrm{e}c^2}
    \left( 1 - \frac{T''}{T_\mathrm{e}}\right),
\end{equation}
where $k$ is the Boltzmann constant, $\sigma_\mathrm{T}$ is the
Thomson scattering cross section, $c$ is the speed of light,
$m_\mathrm{p}$ and $m_\mathrm{e}$ are proton and electron rest masses,
respectively, and $\epsilon$ is the $y$-dependent radiation energy
density. The last terms in Eq.~\ref{eq:lambb} and Eq.~\ref{eq:lambc}
account for the inverse processes, i.e, free-free absorption and
inverse Compton scattering. In this way, heating by free-free
absorption and inverse Compton scattering is included in the cooling
rates. The effective radiation temperatures $T'$ and $T''$ for both
processes are taken to be equal to the photon temperature,
$T_\mathrm{ph}=(\epsilon/a_\mathrm{r})^{1/4}$, where $a_\mathrm{r}$
is the radiation density constant. In this paper, when we discuss
free-free processes (absorption and emission), we mean only thermal
free-free processes, omitting the word ``thermal'' in the following.

The energy balance equation for each spatial layer of the atmosphere
is thus given by
\begin{equation}\label{eq:enbal}
  \Lambda_\mathrm{B} +\Lambda_\mathrm{C} = \begin{cases}
    F_\mathrm{eff}(1-f_\cyc)\rho / {y_0} & \mbox{for $y \leq y_0$,}\\
    0 & \mbox{for $y > y_0$}
    \end{cases}
\end{equation}
where the ``effective flux'' is
\begin{equation}
    F_\mathrm{eff} = \frac{GM_{\ns}\dot{M}}{R_{\ns}}\frac{1}{\pi r_{0}^2},
\end{equation}
and where the term $F_\mathrm{eff}/y_0$ is related to the radiative
energy flux divergence, $\nabla \cdot \vec{F}$, representing the
constant energy release per unit mass. To ensure continuity of the
atmospheric profiles, which is required by the numerical scheme for
the radiative transfer calculation, contrary to ZS69 we choose the
heating to be characterized by an exponential cutoff instead of
Eq.~\ref{eq:enbal},
\begin{equation}\label{eq:enbalexp}
  \Lambda_\mathrm{B} +\Lambda_\mathrm{C} =
  \frac{F_\mathrm{eff}(1-f_\cyc)\rho}{y_0}\,\exp\left(-\frac{y^2}{y_0^2}\right).
\end{equation}
This modification has almost no effect on the obtained solution,
except for smoothing the transition region near $y_0$.

In this part of the modeling, we simplify the model for the
radiative transfer by using the diffusion approximation for the
radiant energy density, $\epsilon(y)$,
\begin{equation}\label{eq:raddif}
    \frac{d\epsilon}{dy}=-\frac{3\varkappa_\mathrm{T}}{c}F(y)
\end{equation}
where $\varkappa_\mathrm{T}=\sigma_\mathrm{T}/m_\mathrm{p}$ is
the opacity of a fully ionized hydrogen plasma due to Thomson
scattering, and $F(y)$ is the flux given by the formula
\begin{equation}\label{eq:difflux}
    F(y) = F_\mathrm{eff}\frac{\left(y-y_0\right)}{y_0}
    \exp\left(-\frac{y^2}{y_0^2}\right).
\end{equation}
Choosing the Marshak boundary condition, i.e., imposing flux
continuity by setting $\epsilon(0)=F_\mathrm{eff}\sqrt{3}/c$,
we can write the solution of Eq.~\ref{eq:raddif} and
Eq.~\ref{eq:difflux} as
\begin{equation}\label{eq:epsilon}
    \epsilon(y) = \frac{3\varkappa_\mathrm{T}}{c}\frac{y_0}{2}F_\mathrm{eff}
    \left(\exp\left(-\frac{y^2}{y_0^2}\right) - 
    \sqrt{\pi}\,\mathrm{erf}\left(\frac{y}{y_0}\right)\right) + C,
\end{equation}
where
\begin{equation}
    C = \sqrt{3}\frac{F_\mathrm{eff}}{c} -
    \frac{3\varkappa_\mathrm{T}}{c}\frac{F_\mathrm{eff}}{2}y_0.
\end{equation}
Finally, hydrostatic equilibrium yields the density distribution
\begin{equation}\label{eq:hydeq}
    P=\frac{2\rho k T_\mathrm{e}}{m_\mathrm{p}} =
    \begin{cases}
        \left(\frac{GM}{R^2}+\frac{\rho_0v_{\mathrm{ff}}^2}{y_0} \right) y,
        & \mbox{for $y \leq y_0$} \\
        \left(\frac{GM}{R^2}+\frac{\rho_0v_{\mathrm{ff}}^2}{y} \right) y,
        & \mbox{for $y > y_0$} \\
    \end{cases}
\end{equation}
where $v_\mathrm{ff}$ is the free fall velocity and where
$\rho_0 = {\dot{M}}/(\pi r_0^2 v_\mathrm{ff})$ is the density of the
falling accretion flow at the upper boundary of the atmosphere.

We obtain the structure of the atmosphere by expressing the density
as a function of the temperature using Eq.~\ref{eq:hydeq} and
substituting it into Eq.~\ref{eq:enbalexp}. The temperature then can
be easily calculated by using a root-finding algorithm, Newton's
method in our case. We emphasize, however, that because of
Eq.~\ref{eq:enbalexp} this approach requires us to assume a value
for the stopping length of the protons, $y_0$. This quantity is not
a free parameter of the model and we will describe our choice of $y_0$
in Sect.~\ref{ssec:calc}, as this assumption is closely related to our
solution of the radiative transfer equation (Sect.~\ref{ssec:radtran}).

\subsection{Polarization and opacities}\label{ssec:sctabs}

Before we describe the transport of the radiation through the
atmosphere (Sect.~\ref{ssec:radtran}), we first need to discuss how
the radiation and photon-electron interactions are affected by the
strong magnetic field. In this section, we address Compton scattering
and free-free absorption in the vicinity of the neutron star
(``magnetic Compton scattering'' and ``magnetic free-free
absorption''), as these are the most relevant processes to be
considered here.

The magnetized plasma of the neutron star atmosphere is an
anisotropic and birefringent medium. It has a strong influence on the
polarization properties of photons, since the interaction between the
plasma and the radiation strongly depends on the frequency and
direction of the photon propagation. Moreover, in strong magnetic
fields the effect of vacuum polarization becomes significant
\citep[see][and \citealt{lai2001} for a review]{adler1971, gnedin1978,
  meszaros1978, meszaros1979}. For the conditions in the accretion
columns of X-ray pulsars this effect can become dominant over plasma
effects. \citet{wang1988} suggested characterizing the relative
importance of plasma and vacuum effects by the ratio of the parameters
$w=(\omega_\mathrm{pe}/\omega)^2$ and
$\delta_\mathrm{V}=(1/45\pi)\alpha(B/B_\mathrm{c})^2$,
\begin{equation}\label{eq:plvac}
    \frac{w}{\delta_\mathrm{V}} = \frac{45\pi}{\alpha}
    \left(\frac{\omega_\mathrm{pe}}{\omega}\right)^2
    \left(\frac{m_\mathrm{e}c^2}{E_\cyc}\right)^2,
\end{equation}
where
\begin{equation}
    \omega_\mathrm{pe}=\sqrt{4\pi n_\mathrm{e}e^2/m_\mathrm{e}}
\end{equation}
is the plasma frequency, $\omega$ is the frequency of the
electromagnetic wave propagating in the medium, $\alpha$ is the
fine structure constant, and
$B_{c}=m_\mathrm{e}^2 c^3 / e \hbar$ is the critical magnetic field
in the quantum-electrodynamical Schwinger limit. For
$w/\delta_\mathrm{V}<1$, the wave propagation is mainly affected
by vacuum polarization. Conversely, when $w/\delta_\mathrm{V}>1$,
the effects of magnetized plasma play a major role. Finally, if
$w/\delta_\mathrm{V}\approx1$, both plasma and vacuum effects have
to be taken into account.

The radiation propagates in the form of two polarization normal modes.
These are defined by the direction of the electrical vector of the
radiation, $\vec{e}$, in relation to the plane formed by the
wave vector of the photon, $\vec{k}$, and the magnetic field
vector, $\vec{B}$. In the ordinary mode (hereafter also referred
to as ``mode~2''), $\vec{e}$ oscillates in the
$(\vec{k},\vec{B})$ plane, while in the extraordinary mode
(``mode~1''), $\vec{e}\perp\left(\vec{k},\vec{B}\right)$.

We mentioned that in the presence of strong magnetic fields, due to
the quantization of the electron motion perpendicular to the magnetic
field on Landau levels photon-electron interactions, such as Compton
scattering and free-free absorption, become resonant processes.
Moreover, these processes proceed differently for each of the
polarization modes. The scattering and absorption cross sections of
both modes differ dramatically in magnitude, especially in the soft
X-ray range, and are strongly energy dependent, which has consequences
for the formation of the continuum as discussed below.

Expressions for the cross sections for scattering and free-free
absorption have been obtained by various authors, using different
approximations as well as different photon polarization modes, such as
pure plasma, pure vacuum, or plasma+vacuum modes \citep[see,
e.g.,][]{ventura1979a, kirk1980, nagel1981b, daugherty1986,
  bussard1986, mushtukov2016, sina1996}. Here, we consider one-photon
Compton scattering, following the quantum mechanical derivations for
differential and total scattering cross sections given by
\citet{nagel1981b}. This derivation neglects the electron spin and
assumes that the initial and final electron occupies the lowest Landau
level, but takes into account thermal electron motion. Thus, these
cross sections are appropriate for a hot, non-relativistic,
$k_\mathrm{B}T_\mathrm{e} \ll m_\mathrm{e} c^2$, plasma.

The probability that a photon with initial energy $E$ and polarization
vector $\vec{e}$, which propagates in the direction $\vec{k}$, will
have the characteristics $E'$, $\vec{k'}$ and $\vec{e'}$ after the
scattering event is given by the differential cross section,
\begin{multline}\label{eq:difcrs}
        \frac{\diff^{2} \sigma}{\diff E' \diff\Omega'}
        (E,\vec{k},\vec{e}\rightarrow E',\vec{k'},\vec{e'}) \\
        =r_\mathrm{e}^{2}\frac{E'}{E\;}\int \diff p f(p)
        |\langle \vec{e'} |\vec{\Pi}| \vec{e} \rangle|^{2}
        \delta\left(E-E'+\frac{p^2}{2m_\mathrm{e}}-\frac{p'^2}{2m_\mathrm{e}}\right),
\end{multline}
where $r_\mathrm{e}$ is the classical electron radius,
$\diff\Omega=\sin{\theta}\diff\theta\diff\phi$ is the solid angle
element, with the azimuthal angle, $\phi$, and the angle $\theta$
between vectors $\vec{k}$ and $\vec{B}$, $f(p)$ is the
non-relativistic Maxwell distribution for the electron momentum $p$,
and $\vec{\Pi}$ is the scattering amplitude
\citep[see Eq.~5 of][]{nagel1981b}. The $\delta$ function expresses
conservation of energy in the scattering event, with $p$ and $p'$
denoting the initial and final momentum of the electron, respectively.
From the optical theorem the total
cross section is then
\begin{equation}\label{eq:crs}
    \sigma = -4\pi r_\mathrm{e}\frac{c\hbar}{E}\mathrm{Im}\,\hat{\vec{\Pi}},
\end{equation}
where the matrix elements of the scattering amplitude, $\hat{\vec{\Pi}}$,
are averaged over the Maxwellian distribution.

We assume pseudo-local thermodynamic equilibrium and treat cyclotron
absorption and reemission as scattering, assuming that the collisional
de-excitation rate is lower than the radiative decay rate, consistent
with our atmosphere model and the classical approach of
\citet{nagel1980}, \citet{meszaros1985a}, and \citet{alexander1989}.
Under these assumptions, the thermal production rate of photons is
given by $j=\alpha^\mathrm{ff}B(T_{\mathrm{e}})$, where
$B(T_\mathrm{e})$ is the Planck function of the local electron
temperature and where the magnetic free-free absorption coefficient,
$\alpha^\mathrm{ff}$, includes the continuum and the resonant part
due to cyclotron processes \citep[see, e.g., Eq.~7 of][]{meszaros1985a}.

Although the \citet{nagel1981b} cross sections are given for plasma
normal modes, they allow an easy replacement of the polarization
vectors with vacuum ones (see, e.g., Eq.~\ref{eq:difcrs}). For the
photon energy range used in the calculations and throughout most of
the atmosphere, we find with Eq.~\ref{eq:plvac} that the atmosphere is
dominated by vacuum effects. In the following, we therefore use the
pure vacuum normal modes to describe the photon polarization state, as
given by \citet[][their Eq.~12a--f]{wang1988}. This approach results
in a substantial acceleration of the computations. The other reason to
avoid mixed plasma and vacuum normal modes is the ambiguity of the
definition of the modes, as well as the fact that they become
non-orthogonal when both vacuum and plasma effects are taken into
account \citep{soffel1983}. This ambiguity would require a more
careful investigation \citep{ho2003} and is left for a future
development of the model. 

\begin{figure}
    \resizebox{\hsize}{!}{\includegraphics{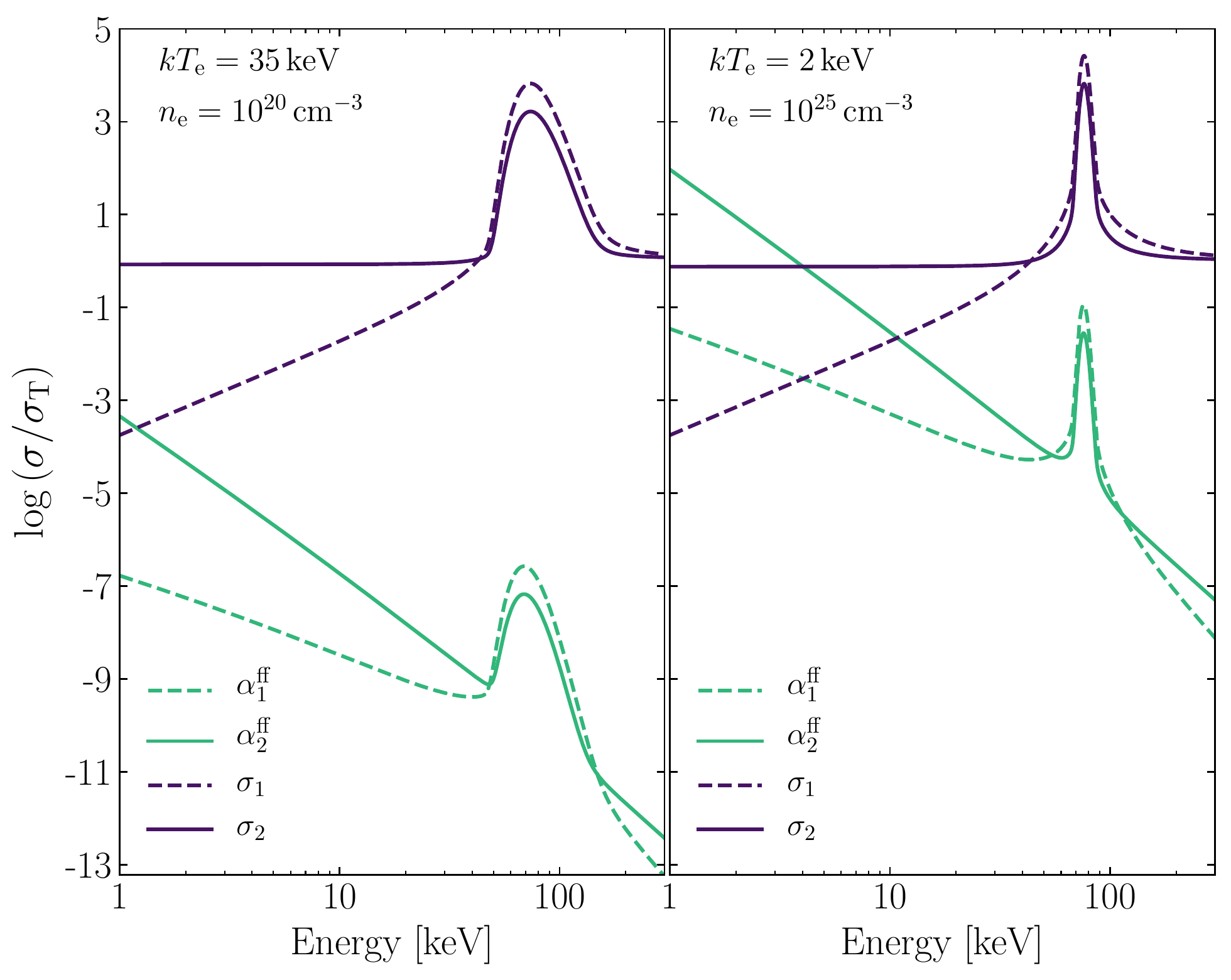}}
    \caption{Total cross section for magnetic Compton scattering
      (purple lines) and the free-free absorption coefficient (cyan
      lines) for the ordinary (solid) and extraordinary (dashed)
      polarization mode, taking into account vacuum polarization.
      \emph{Left:} coefficients for a tenuous and hot plasma with
      $n_\mathrm{e}=10^{20}\,\mathrm{cm^{-3}}$ and
      $kT_\mathrm{e}=35\kev$, as expected for the upper, overheated
      layers of the atmosphere. \emph{Right:} Same but for cold and
      dense plasma with $n_\mathrm{e}=10^{25}\,\mathrm{cm^{-3}}$ and
      $kT_\mathrm{e}=2\kev$, corresponding to the bottom part of the
      atmosphere. The cyclotron line energy is
      $E_\cyc=75\kev$ and the incident photon has a direction of
      $\theta=60^\circ$ with respect to the magnetic field, for both
      cases. }
    \label{fig:crs}
\end{figure}

To illustrate the behavior of the vacuum cross sections, in
Fig.~\ref{fig:crs} we show the magnetic Compton scattering and
magnetic free-free absorption cross sections for ordinary and
extraordinary modes, for two different electron temperatures,
$T_\mathrm{e}$, and electron number densities, $n_\mathrm{e}$,
choosing values that are typical for the atmospheric conditions
relevant to the present study.

\subsection{Polarized radiative transfer}\label{ssec:radtran}

To describe the propagation of the radiation field through the
atmosphere, we need to solve the radiative transfer equation. In this
section, we discuss the transfer equation (taking into account the
azimuthal symmetry of the task), the boundary conditions of the
radiative transfer, and its numerical solution.

\subsubsection{Transfer equation}

In the case of the emitting region considered here, we can simplify
the treatment of radiative transfer by assuming that the atmosphere is
a semi-infinite, plane-parallel slab. For such medium we can write two
equations that govern the radiative transfer in the outgoing, and
incoming directions,
\begin{equation}\label{eq:rtrans0}
    \pm\mu\frac{\partial I_\pm}{\partial z} = \kappa(S - I_\pm),
\end{equation}
where $\mu=\cos{\theta}$ is restricted to the half-range $[0,1]$,
$I_\pm=I(E, \pm\mu, z)$ is the specific intensity,
$\kappa = \sigma + \alpha^\mathrm{ff}$ is the total opacity, and $S$
is a source function.

By introducing symmetric and anti-symmetric 
averages of the specific intensity,
\begin{subequations}\label{eq:uvfunc}
  \begin{align}
    u &= \frac{1}{2}(I_{+} + I_{-})\label{eq:ufunc}\\
    v &= \frac{1}{2}(I_{+} - I_{-})\label{eq:vfunc},
  \end{align}
\end{subequations}
\citet{feautrier1964} suggested an approach that allows
presenting the radiative transfer equation as a second-order
differential equation. Here, $u$ and $v$ are
the analogs of zero- and first-order moments of the specific
intensity, respectively. By adding and subtracting two equations
of Eq.~\ref{eq:rtrans0}, we obtain
\begin{subequations}
  \begin{align}
    \mu\frac{\partial v}{\partial z} &= \kappa(S - u) \label{eq:veq} \\
    \mu\frac{\partial u}{\partial z} &= - \kappa v \label{eq:ueq}.
  \end{align}
\end{subequations}
We can now obtain a second-order equation by expressing $v$ in
terms of $u$ from Eq.~\ref{eq:ueq} and substituting it to
Eq.~\ref{eq:veq}, which yields
\begin{equation}\label{eq:rtrans1}
    \frac{\mu^2}{\kappa}\frac{\partial^2 u}{\partial z^2} = \kappa(u-S).
\end{equation}
In our case, the source function is given by
\begin{multline}\label{eq:sourcef}
    S = \frac{4\pi}{\kappa}\!\int_{-\infty}^\infty\!\diff E'\!
    \int_0^1\!\diff\mu'\!\frac{\diff^2\sigma}{\diff E' \diff\mu'}
    \left(E, \mu \rightarrow E', \mu'\right)\frac{E}{E'}u'\\
    + \frac{\alpha^\mathrm{ff}}{\kappa}\tilde{u}_\mathrm{B} + S_\cyc,
\end{multline}
where the integrals represent electron scattering with partial energy
and angular redistribution, with $u'=u(E',\mu',z)$, and where we took
into account the azimuthal symmetry of the problem. The second term
describes a contribution from thermal absorption and emission with
$\tilde{u}_\mathrm{B}$ being the equilibrium spectrum, which is
Planckian with the local electron temperature. The
source of non-thermal cyclotron emission at each layer of the
atmosphere, $S_{\cyc}$, is given by
\begin{equation}\label{eq:cycsource}
  S_\cyc(E) = F_\mathrm{eff}
  \frac{f_{\cyc}}{4\pi} \frac{1}{y_0}
  V_{\mathrm{D}}(E, T_{\mathrm{e}})\exp\left(-\frac{y}{y_0}\right),
\end{equation}
where the profile $V_{\mathrm{D}}$ is determined by the Doppler
broadening
\begin{equation}
  V_{\mathrm{D}} = \frac{1}{\sqrt{\pi}\Delta E_{\mathrm{D}}}
  \exp\left(-\frac{(E-E_{\cyc})^2}{\Delta E_{\mathrm{D}}^2}\right),
\end{equation}
with
\begin{equation}
  \Delta E_{\mathrm{D}}=E_{\cyc}\sqrt{2kT_{\mathrm{e}}/m_{\mathrm{e}}c^2}.
\end{equation}
We note that Eq.~\ref{eq:rtrans1}, together with Eq.~\ref{eq:sourcef},
does not include stimulated processes directly (the stimulated
scattering term is omitted), such that the problem is linear. However,
ignoring stimulated scattering for optically thick plasma can lead to
a significant excess of photons (${>}50\%$) at high energies
\citep{alexander1989}. \citet{meszaros1989} introduced a compromise
procedure that allows treating the stimulated scattering while
avoiding the nonlinear problem by assuming an artificial detailed
balance relation,
\begin{multline}\label{eq:detbal}
  \frac{\diff^{2} \sigma}{\diff E'\diff\mu'}(E,\mu\rightarrow
  E',\mu')\\
  =\left(\frac{E'}{E}\right)^{2}
  \left(\frac{\exp\left(E/kT_{\mathrm{e}}\right)-1}
  {\exp\left(E'/kT_{\mathrm{e}}\right)-1}\right)
  \frac{\diff^{2} \sigma}{\diff E \diff\mu}(E',\mu'\rightarrow E,\mu).
\end{multline}
In equilibrium, this expression leads again to a Planckian photon
spectrum. It fully reproduces the low energy part of the spectrum
obtained by solving the nonlinear problem and slightly
underestimates the high energy photon flux while keeping the same
spectral shape. \citep[see Figs.~2,3 of][for a direct comparison with
the nonlinear problem]{alexander1989}.

We next consider polarization effects. These require us to separate
the radiation field into the ordinary and extraordinary modes, which
are coupled through the scattering integral. In order to remain
consistent with the description of the atmospheric structure, we
rewrite the transfer equation in terms of the column density $y$, by
taking into account that $dy=-\rho dz$, as
\begin{multline}\label{eq:rtrans}
  \frac{\mu^2}{\kappa_p/\rho^2}\frac{\partial^{2}
    u_p} {\partial y^{2}} = \kappa_p u_p \\
  -4\pi\sum\limits_{p'=1, 2}\iint \frac{\diff^{2} \sigma_{pp^{\prime}}}
  {\diff E^{\prime} \diff\mu'} \frac{E}{E^{\prime}} u_{p'} \,
  \diff E' \diff\mu' -
  \alpha_p^\mathrm{ff}\frac{\tilde{u}_\mathrm{B}}{2}-\kappa_p\frac{S_\cyc}{2},
\end{multline}
where the subscript $p=1$ for the extraordinary mode and $p=2$ for the
ordinary mode, $u_p=u_p(E, \mu, y)$, the opacity
$\kappa_p=\sigma_p + \alpha_p^\mathrm{ff}$, and the factor $\tfrac{1}{2}$
occurs from the separation of the two modes.

\subsubsection{Boundary conditions}

Finally, in order to solve Eq.~\ref{eq:rtrans}, we need to specify the
boundary conditions at the top of the atmosphere, $y=0$, (``upper''
boundary) and at $y=y_\mathrm{max}$ (``lower'' boundary). At the upper
boundary we assume that there is no external illumination of the slab,
$I_{-}(0)=0$. In this case, from Eq.~\ref{eq:uvfunc} we find $u(0)=v(0)$
and Eq.~\ref{eq:ueq} yields
\begin{equation}\label{eq:bctop}
  \frac{\mu}{\kappa_p/\rho} \, \frac{\partial u_p}{\partial y} = -u_p.
\end{equation}
The assumption of free photons emerging from the upper boundary is
consistent with the physical picture of a low accretion rate, where
the effects of the radiative transport in the moving plasma above the
polar cap can be ignored.

At the lower boundary of the atmosphere, at high optical depths
corresponding to the chosen $y_\mathrm{max}=10^3\gcm$ (and high
densities, as will be shown in Sect.~\ref{res:atm}), the radiation
field can be treated in the diffusion approximation, i.e.,
$I(y_\mathrm{max})=B(y_\mathrm{max}) + \mu(\kappa^{-1}
\left|\partial{B}/\partial{y}\right|)_{y_\mathrm{max}}$ \citep[see,
e.g,][chap.~2]{mihalas1978}. As the mean intensity
$u(y_\mathrm{max})=B(y_\mathrm{max})$, from Eq.~\ref{eq:uvfunc} we
obtain
$v(y_\mathrm{max})=\mu(\kappa^{-1}
\left|\partial{B}/\partial{y}\right|)_{y_\mathrm{max}}$. In our 
case, Eq.~\ref{eq:enbalexp} leads to the existence of an extended
cold isothermal layer below $y_0$ (see Sect.~\ref{res:atm}), and
therefore Eq.~\ref{eq:ueq} yields
\begin{equation}\label{eq:bcbot}
    \frac{\partial u_p}{\partial y} = 0.
\end{equation}
These boundary conditions together provide a self-emitting slab, where
free-free emission, including both continuum and resonant parts of the
coefficient, and non-thermal cyclotron emission are the only sources
of photons.

\subsubsection{Numerical solution}

We model the radiative transfer by discretizing Eq.~\ref{eq:rtrans}
for each polarization mode using an appropriate angular \{$\mu$\},
energy \{$E$\}, and column density (``spatial'') \{$y$\} grid. We then
write a matrix equation for each $y$-layer and solve the system of
matrix equations using a tridiagonal matrix algorithm. This approach to
solving the transfer equation for atmospheres with predefined
temperature and density profiles is known as the ``Feautrier method''
\citep{feautrier1964}. The method is efficient even for optically
thick atmospheres and allows for partial angular and energy
redistribution. The application of the method to the polarized
radiation in magnetized atmospheres was described by
\citet{nagel1981b}, \citet{meszaros1985a} \citep[see also][chap.~6, for
the classical formulation]{mihalas1978}. We follow the prescriptions
of \citep{auer1967} to improve the boundary conditions by performing a
Tailor series expansion away from the boundary, including terms up to
the second order. Our code for the radiative transfer simulation,
\texttt{FINRAD}, is written in Python 3 and based on sparse matrices.
It was extensively tested using known analytical and numerical
solutions for scattering atmospheres and the well-known transfer code
$\texttt{XILLVER}$ for reflection from accretion disks
\citetext{J.~A.~Garcia\ priv.\ comm.}. Some examples of this code
verification are presented in Appendix~\ref{app:tests}.

\subsection{Complete simulation}\label{ssec:calc}

In order to determine the structure of the atmosphere and then use it
to solve the radiative transfer equation, we first need to obtain an
estimate of the stopping length of protons, $y_0$. The detailed
modeling of the proton stopping at the level of, e.g.,
\citet{kirk1982} or \citet{miller1987}, is outside the scope of this
paper. Instead, we iteratively find $y_0$ for each given set of the
parameters, \{$\dot{M}$, $r_0$, $E_\cyc$, $f_{\cyc}$\}, by requiring
that energy is conserved inside the emitting region. Specifically, for
a chosen $\dot{M}$, we require that all kinetic energy of the accretion
flow is converted to the radiation emerging from the atmosphere 
\begin{equation}
L_\mathrm{acc} = \frac{1}{2}\dot{M}v_\mathrm{ff} =
\frac{GM_\ns\dot{M}}{R_\ns}.
\end{equation}
We restrict the calculated bolometric flux by setting
\begin{equation}\label{eq:encons}
L_\mathrm{num} = \pi r^2_0  \int (F_1(E)+F_2(E)) \diff E
\stackrel{!}{=} L_\mathrm{acc},
\end{equation}
where $F_1$ and $F_2$ are the specific fluxes of the extraordinary and
the ordinary mode, respectively.
\begin{figure}
    \resizebox{\hsize}{!}{\includegraphics{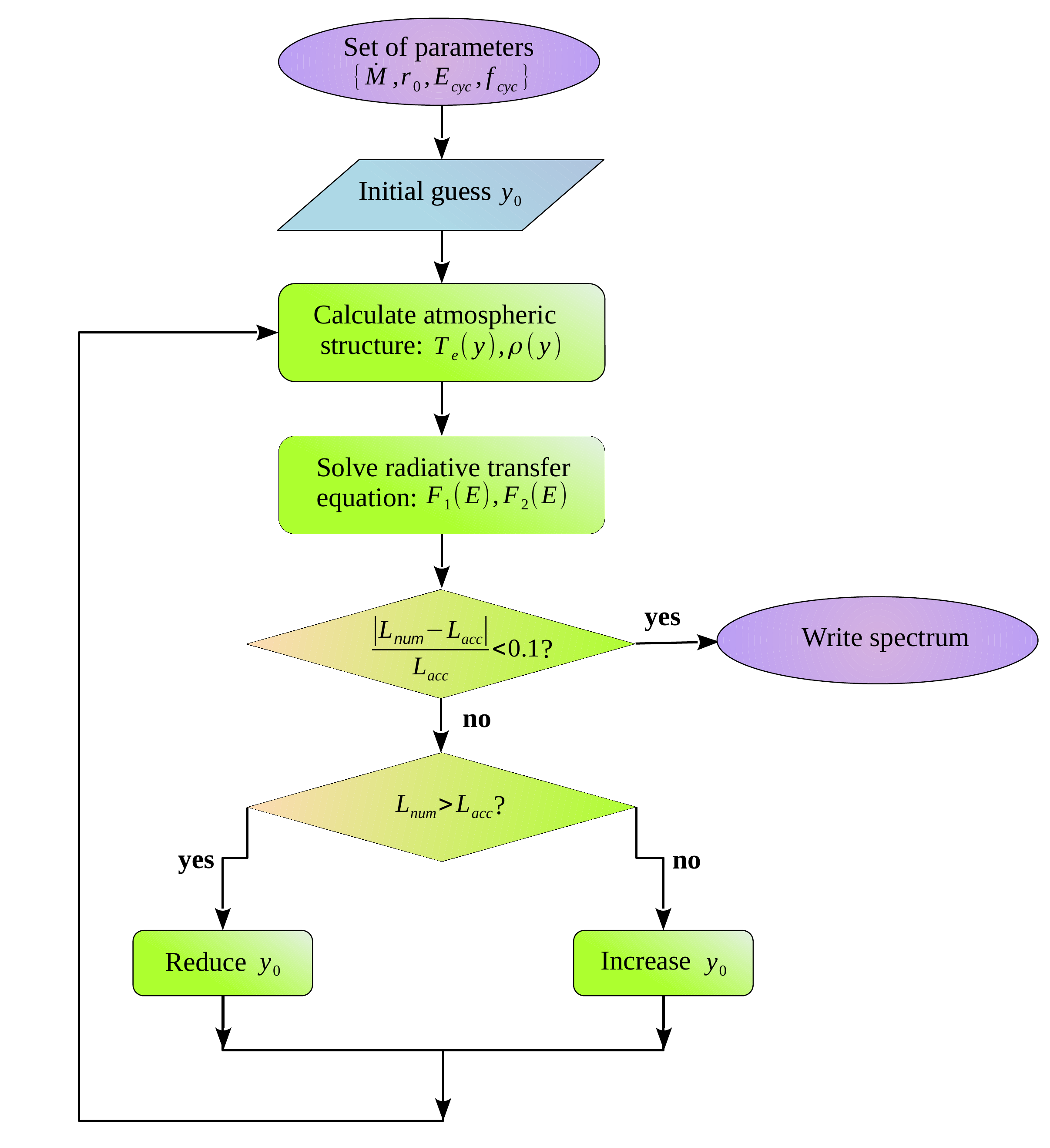}}
    \caption{Flowchart of the complete numerical solution.}
    \label{fig:chart}
\end{figure}

In order to obtain a self-consistent model of the atmosphere, for each
set of parameters \{$\dot{M}$, $r_0$, $E_\cyc$, $f_\cyc$\} we choose
an initial value for $y_0$ (usually, $y_0=6$--$9\gcm$), calculate the
atmospheric structure based on the discussion in
Sect.~\ref{ssec:atm}, and then solve the radiative transport as
described in Sect.~\ref{ssec:radtran} to obtain the flux. We then
compare the resulting luminosity of the slab, $L_\mathrm{num}$, to
$L_\mathrm{acc}$ and, if necessary, adjust $y_0$ iteratively until
energy conservation is achieved at better than 10\% tolerance, i.e.,
until $|L_\mathrm{num}-L_\mathrm{acc}|/L_\mathrm{acc}<0.1$. The total
process of the simulation is illustrated in Fig.~\ref{fig:chart}.

For the discretization of the spatial scale, we use a logarithmic grid
within the range $y=10^{-3}\mbox{--}10^{3}\gcm$. We continue our
grid to such high column densities (compared to $y_0$) to ensure
that the medium has a sufficiently high optical depth for both
polarization modes over a wide energy range. Angular discretization
is performed according to the 8-point double-Gauss formula for $\mu$
given by the roots of the Legendre polynomial on the range $[0, 1]$
\citep[see][chap.~3]{sykes1951, mihalas1978}. The energy grid
requires more care, as it is necessary to ensure having a fine
grid around the cyclotron resonance while the grid in the
continuum can be coarser. At the same time, however, the steps
in the energy grid should not change abruptly as this can lead
to instabilities in the numerical scheme and also strongly
affects the redistribution in energy. Our compromise solution
is to use a non-uniform grid generator where the points condense
gradually towards the so-called ``target'', which in our case is
the cyclotron resonance. Here, we apply the generator by
\citet{tavella2000}, which gives the grid points
\begin{equation}\label{eq:engrid}
    E_{i}=E_{\cyc} + \zeta\sinh\left( c_2\frac{i}{N} +
    c_1\left(1-\frac{i}{N}\right)\right),
\end{equation}
where
\begin{equation}
  c_{1,2} = \sinh^{-1}\left( \frac{E_{\mathrm{min},\mathrm{max}} - E_{\cyc}}{\zeta}\right),
\end{equation}
and where the parameter $\zeta$ determines the uniformity of the
grid. In our calculations, we choose $\zeta=0.3$ and set
$E_{\mathrm{min}}=0.8\kev$, $E_{\mathrm{max}}=250\kev$, with a
total of 300 points on the energy grid.

\section{Spectral formation and atmospheric structure}\label{sec:results}

In this section, we describe the obtained atmospheric structure,
the formation of the emergent radiation field, and its spectral
shape. We base our discussion on a parametric study of a set of 144
spectra in which we varied the mass accretion rate, polar cap radius,
magnetic field, and the fraction of accretion energy going into
non-thermal collisional excitation of electrons over the following
parameter ranges:
$\dot{M}=[4\mbox{--}16]\times10^{13}\gsec$,
$r_0=[80\mbox{--}140]\,\mathrm{m}$,
$E_\cyc=[50\mbox{--}80]\kev$,
$f_\cyc=[0\mbox{--}0.4]$. The neutron star mass and radius are fixed
to $1.4\msun$ and $12\,\mathrm{km}$, respectively.

\subsection{Atmospheric structure}\label{res:atm}

\begin{figure*}
    \centering
    \includegraphics[width=\textwidth]{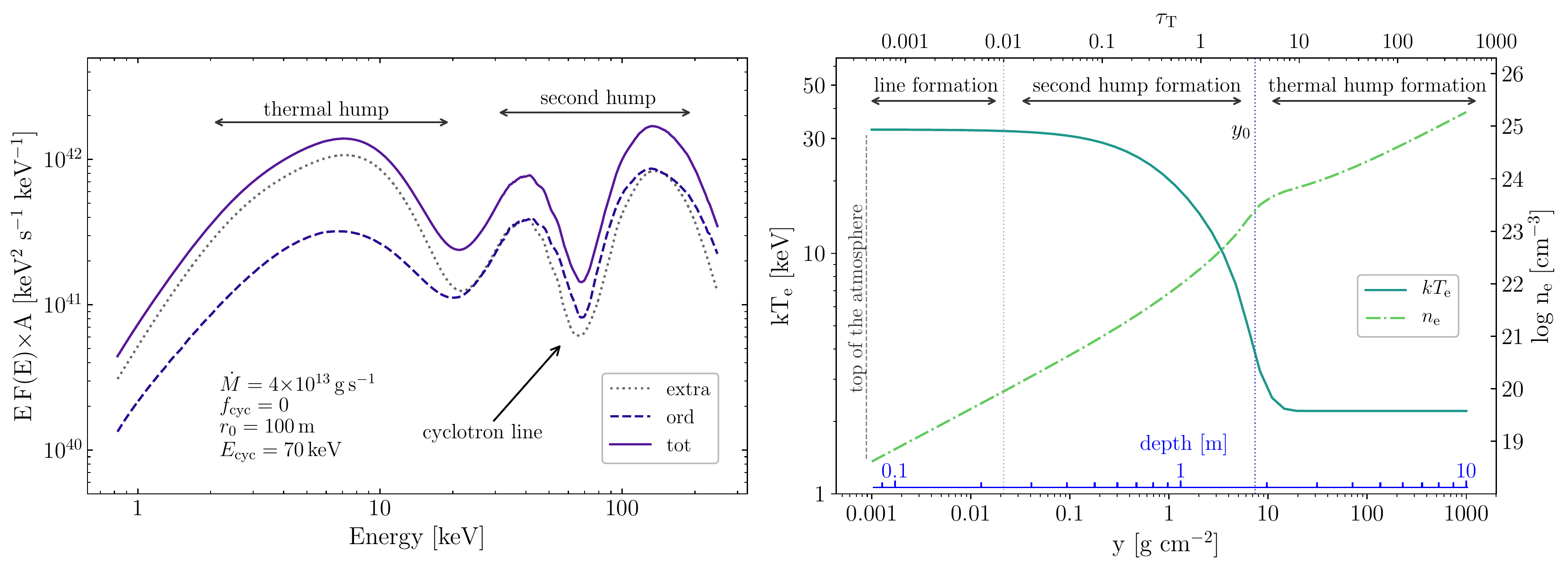}
    \caption{\emph{Left:} Emerging spectrum in two polarization modes.
             Solid lines describe the total flux, dotted lines extraordinary,
             and dashed lines ordinary photons. \emph{Right:} Atmospheric structure: electron temperature (solid) and number
             density (dash-dotted). The major spectral components and
             associated formation regions are indicated in both panels.}
    \label{fig:specatm}
\end{figure*}

We start our discussion with a description of the internal structure
of the atmosphere, which to a large extent determines the shape of the
emergent spectrum. Figure~\ref{fig:specatm}(right) shows the electron
temperature and density distribution along the vertical axis in the
atmosphere for a typical set of parameters. The atmosphere has two
principal parts determined by $y_0$. In the upper part, where
$y < y_0$, the heating by accreted matter significantly affects the
energy balance (Eq.~\ref{eq:enbalexp}). The temperature of this region
mainly depends on $y_0$ and reaches values of $30\mbox{--}40$\kev.
Media with shorter $y_0$ have a higher temperature as the accretion
energy dissipates in a smaller volume. In the dense lower part of
the atmosphere, only the balance of free-free processes and Compton
scattering determines the low equilibrium temperature (${\sim}2\kev$).
A constant level of the local electron temperature is reached soon
after $y_0$ and depends mainly on the ratio $\dot{M}/\pi r_0^2$, which
controls the properties of the falling flux at the upper border of
the atmosphere through the continuity equation. We also emphasize
which parts of the atmosphere play a major role in the formation of
individual spectral components. We will focus on this in the following
sections.

In addition to the column density, Fig.~\ref{fig:specatm}(right)
shows the scales for optical and geometrical depths in the atmosphere.
We emphasize that even
though the total column density $y_\mathrm{max}$ is high,
the atmosphere remains a geometrically thin layer of only 
${\sim}10\,\mathrm{m}$. It justifies our assumption about a
constant magnetic field along the vertical direction inside the
atmosphere and allows us to estimate the effect of gravitational
redshift on the emitted spectrum (see, for example,
Sect.~\ref{sec:gx304-1}). Besides, it validates the choice of
one-dimensional treatment for the radiative
transfer, as the surface area of the sidewall of the atmospheric 
slab, $A_\mathrm{sw}=2\pi z_\mathrm{max}r_0$ is much smaller than
the polar cap area, $A=\pi r_0^2$. The simple estimate
on the corresponding luminosities yields
$L_\mathrm{sw}/L_\mathrm{top} < 2z_\mathrm{max}r_0\ll1$.
Thus, the radiation flux through the sides can be neglected. Of
course, the escape of photons from the sidewall is by no means free
and is determined by the radiant thermal conductivity of the
magnetized plasma.

\subsection{Formation of the emitted spectrum}

We can now discuss the spectral formation in the atmosphere with
pre-calculated structure as a result of photon-electron interactions
in a highly magnetized medium in detail. We first investigate the
formation of the spectrum in the absence of non-thermal collisional
excitations in the atmosphere, i.e., $f_\cyc=0$. This case corresponds
to a very strong magnetic field where the energy of the flow is
insufficient to excite the electrons even to the first Landau level.
This approach allows us to illustrate the details of the second hump
origination.

A typical energy flux (times energy) for low mass accretion rates
onto a magnetized neutron star from the \polcap model is shown in
Fig.~\ref{fig:specatm}(left), corresponding
to the atmospheric structure from Fig.~\ref{fig:specatm}(right). The
overall emerging spectrum has two dominant spectral components: a
softer, thermal hump and a high-energy (``second'') hump characterized
by a cyclotron line. 

\begin{figure}
    \resizebox{\hsize}{!}{\includegraphics{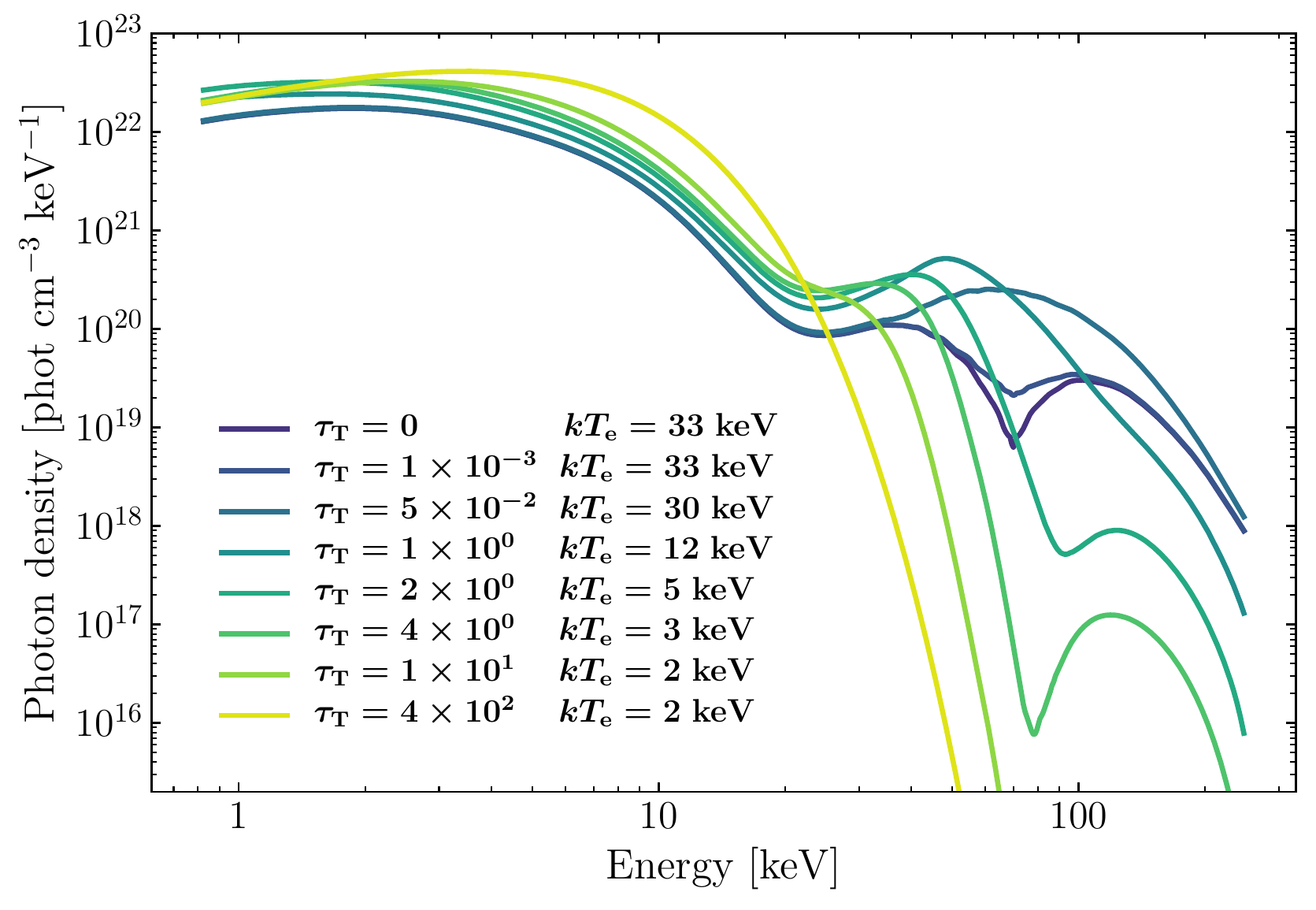}}
    \resizebox{\hsize}{!}{\includegraphics{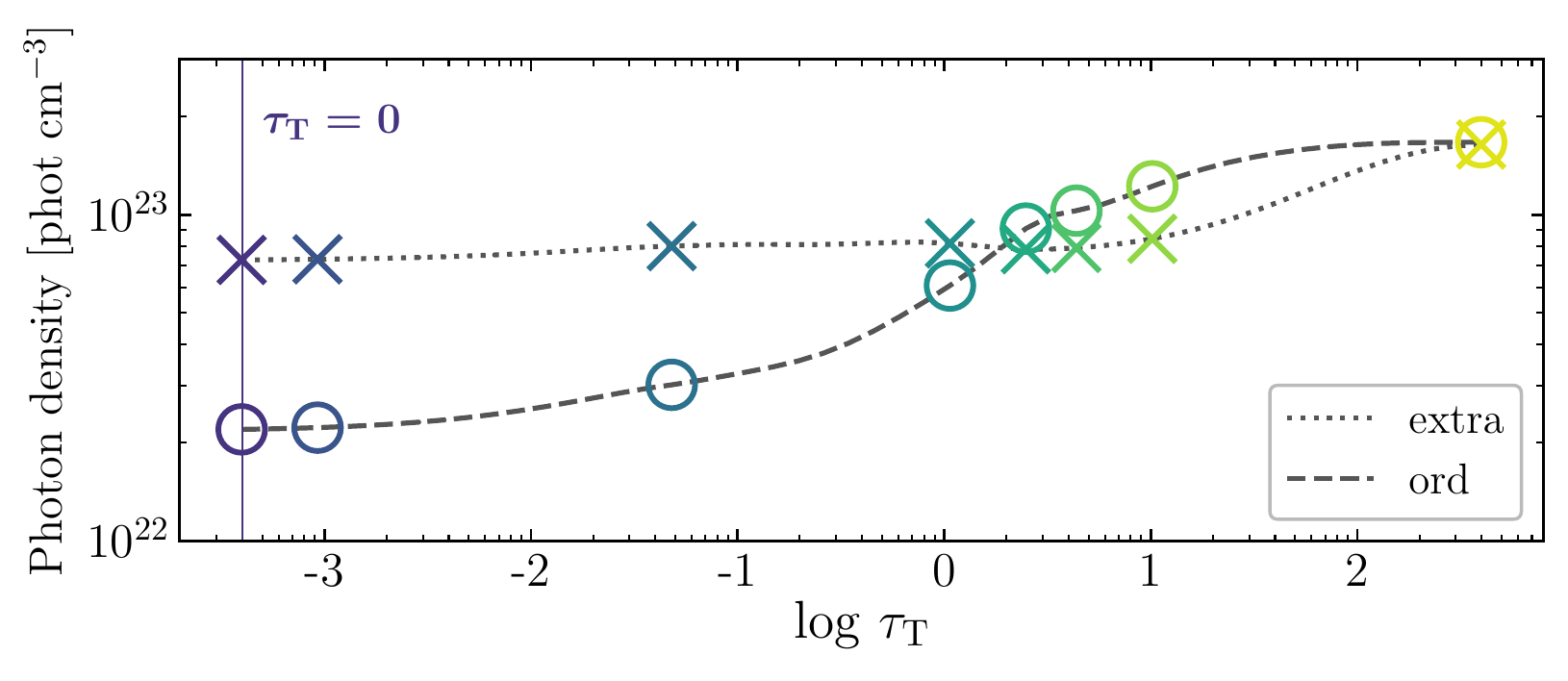}}
    \caption{\emph{Top:} Photon density at different optical depths of
             the neutron star atmosphere, $\tau_\mathrm{T}$, summed
             over polarization modes. The set of model parameters is the
             same as in Fig.~\ref{fig:specatm}. \emph{Bottom:} Photon
             density integrated over photon energies in each polarization
             mode. Markers correspond to the lines of a similar color in
             the top panel; circles indicates ordinary and crosses
             extraordinary photons.}
    \label{fig:fluxspat}
\end{figure}

We can understand the formation of the spectrum by studying how
the photon density
\begin{equation}\label{eq:phdens}
    f_p(E, y)=\frac{4\pi}{c}\frac{J_p(E, y)}{E}
\end{equation}
changes throughout the atmosphere in our model. The mean intensity
$J$ can be found at each internal spatial layer by calculating
the quadrature over angles, $J_p(E, y)=\sum \varw_\mu u_p(E, \mu, y)$,
where $w_\mu$ are the weights given by double-Gauss formula.
The top part of Fig.~\ref{fig:fluxspat} shows the evolution of the
photon density spectrum inside the atmosphere. Starting at
the bottom of the slab, at the highest optical depth,
$\tau_\mathrm{T}\approx4\times10^{2}$,
the free-free absorption dominates opacities for both modes (see the
right-hand side of the Fig.~\ref{fig:crs}, displaying the scattering
cross sections for similar conditions) and the radiation field is at
local thermodynamic equilibrium. The photon spectrum is Planckian
corresponding to the local electron temperature $kT_\mathrm{e} = 2\kev$.
At higher, but still optically thick, layers of the slab, one can see
the effect of Compton up-scattering and the formation of the cyclotron
scattering feature. Since the scattering coefficient in the line core 
is very high, the photons are mainly scattered out of the core, forming
pronounced shoulders to the line on top of the Comptonized continuum.
Closer to $\tau_\mathrm{T}\approx 1$, the temperature profile is
affected by the heating of the slab by accreted matter. It results
in more efficient Compton up-scattering of softer photons and
further diffusion of the photons away from the line core, which
together with the significant temperature gradient fills up the
feature, making it almost completely disappear. At small optical
depths, $\tau_\mathrm{T}\lesssim 10^{-2}$, the continuum is not
further affected by scattering, except for the region near the
cyclotron energy. As the upper part of the atmosphere has a very
high electron temperature, the resulting cyclotron line is very
broad. We find that its width can be well estimated by assuming
pure Doppler broadening \citep{meszaros1985a}
\begin{equation}\label{eq:dopp}
    E_\mathrm{FWHM} \approx E_\cyc 
    \left( 8\ln 2 \frac{kT_\mathrm{e}}{m_\mathrm{e}c^2}\right)^{1/2}\langle\cos{\theta}\rangle .
\end{equation}
In the case discussed here, $kT_\mathrm{e}=33\kev$, $E_\cyc=70\kev$,
and $\langle\cos{\theta}\rangle=2/\pi$ result in
$E_\mathrm{FWHM} \approx 27\kev$, in agreement with the line
width from our simulation.

The bottom panel of Fig.~\ref{fig:fluxspat} shows the energy
integrated photon density profile inside the atmosphere. Deep
in the atmosphere, both photon modes are thermal and mixed. As a
result, radiation is unpolarized. The extraordinary mode
saturates to the equilibrium value much deeper in the atmosphere
due to much lower opacity. Since the mean free path, $\lambda$,
scales as $\lambda\propto\sigma^{-1}$, at low energies
extraordinary photons have a much longer mean free path than
ordinary ones (see Fig.~\ref{fig:crs}). For this reason, soft
extraordinary photons can leave the medium from deeper layers
of the atmosphere, and the radiation becomes polarized. The density
of extraordinary photons is additionally increased by mode
conversion during scattering events. Once an ordinary photon
changed its polarization during scattering, it can leave the
medium. As a result, the spectrum of the emergent radiation is
dominated by soft extraordinary photons. The high-energy photons
and the ordinary mode are trapped in the atmosphere longer by
the resonant scattering. For both polarizations, photon densities
decrease towards the top of the atmosphere, reaching the limiting
values at an optical depth comparable with their mean free path.
We note that these conclusions are in agreement with the results
of \citet{meszaros1985b} for a deep optically thick atmosphere.

In summary, the high-energy component is mainly produced by
resonant magnetic Comptonization in the heated non-isothermal
part of the atmosphere, and then modified by the formation of
the cyclotron line in the optically thin layers. Our polarization
dependent computations also show that the fluxes of ordinary
and extraordinary photons emerging from the atmosphere at the
high-energy component is very similar to each other. This is
caused by the large cross sections around the cyclotron resonance
for both modes. However, more actively comptonized ordinary
photons dominate in the line core.

In contrast, the low-energy component is mainly formed by
extraordinary photons. As extraordinary photons can escape from
deeper layers of the atmosphere, the soft thermal emission in this
mode is not modified dramatically by scattering. For the ordinary
photons, on the other hand, it is significantly changed by Compton
up-scatterings. Here, we see a flattened spectrum in low energy
part of the ordinary mode, which results from saturated
Comptonization, in agreement with previous results for high
optical depth for magnetic \citep{meszaros1985a} and
non-magnetic \citep{sunyaev1980} atmospheres. Considering this and
the dominant contribution of the extraordinary mode to the low
energy part of the spectrum, we can therefore expect the outgoing
radiation to have a high degree of polarization. The distinguished
contribution of the ordinary and extraordinary photons to the soft
and hard energy band of the emerging spectrum is also well known
from the studies of \citet{ceccobello2014} and
\citet{lyubarskii1988a,lyubarskii1988b}.

\subsection{Parameter dependency of the emitted spectrum}\label{sec:pardep}

\begin{figure*}
    \centering
    \includegraphics[width=17cm]{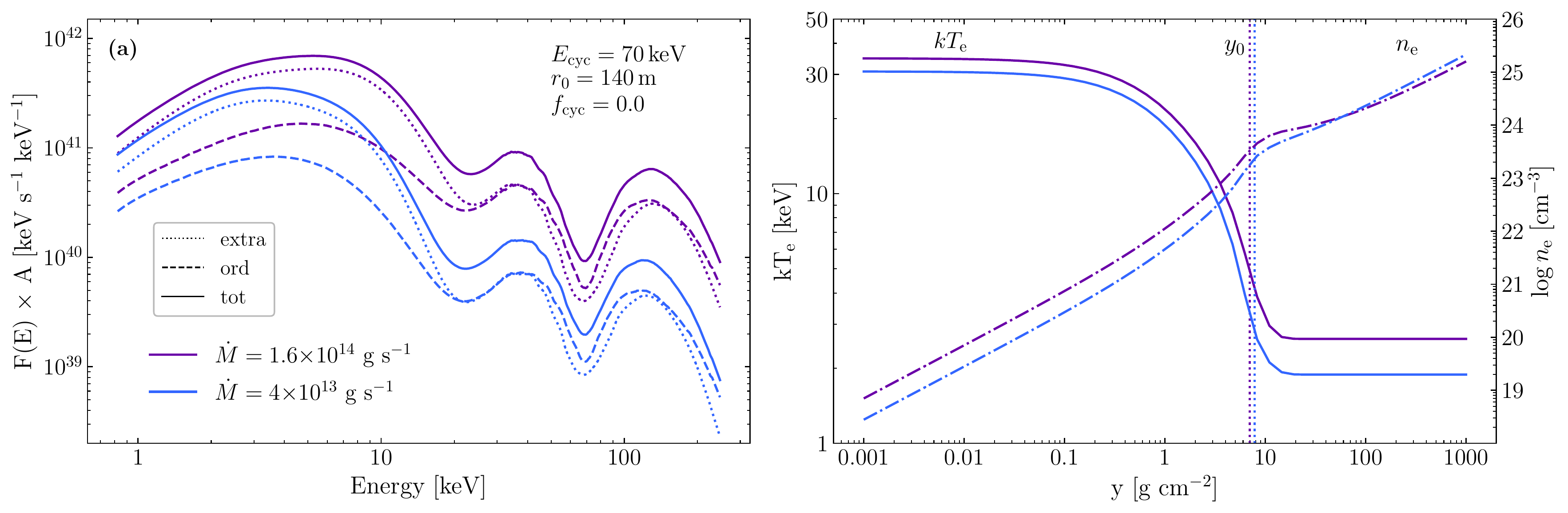}\\
    \includegraphics[width=17cm]{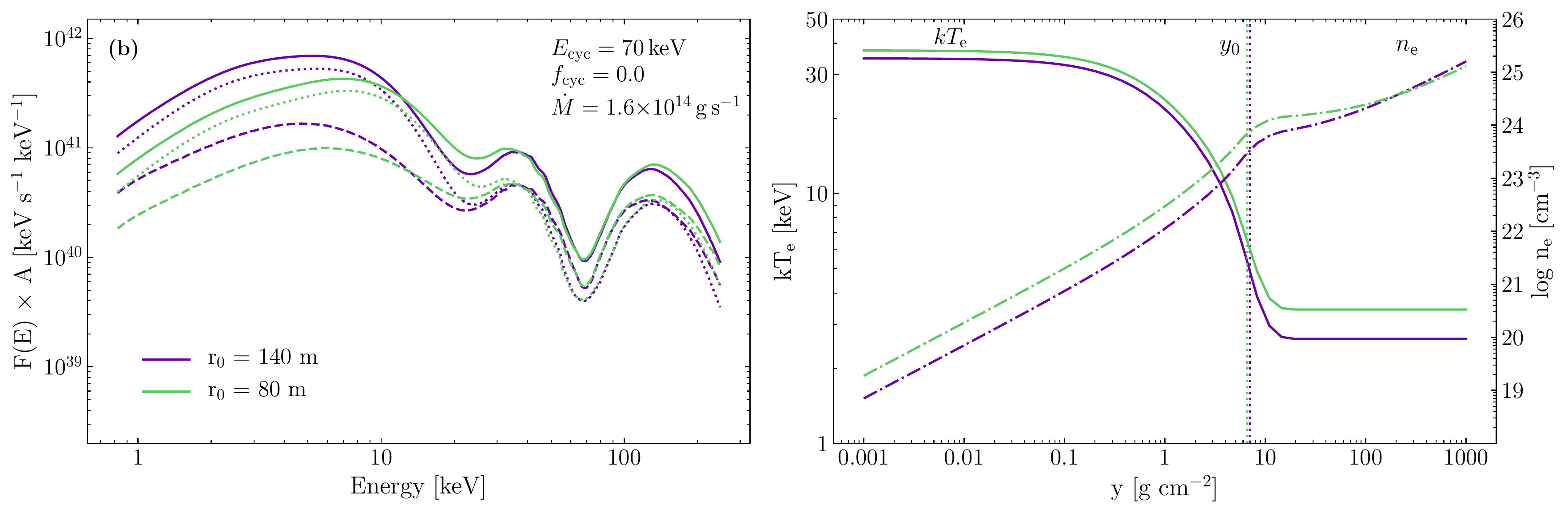}\\
    \includegraphics[width=17cm]{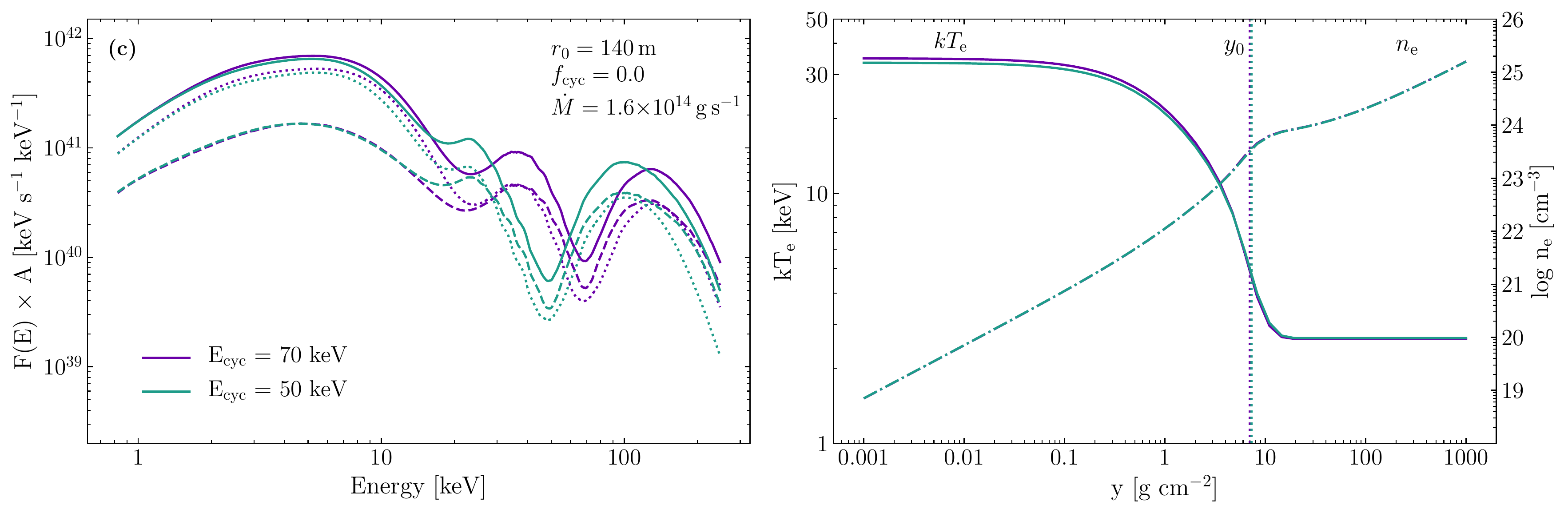}\\
    \includegraphics[width=17cm]{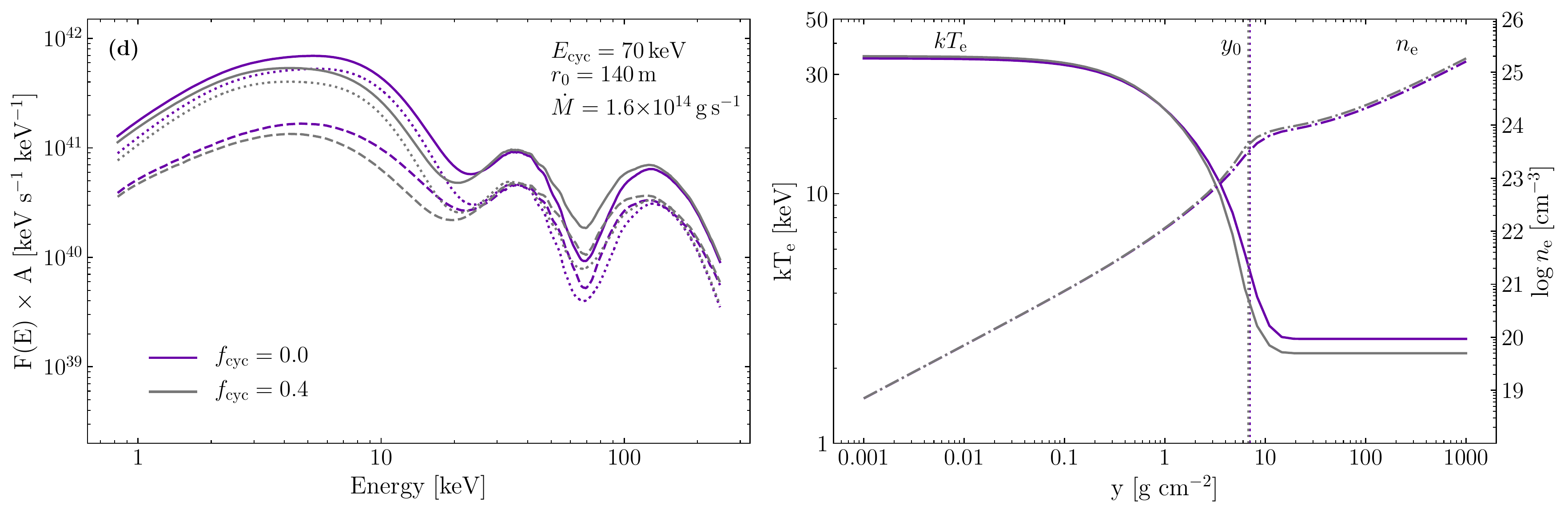}
        
    \caption{Dependency of the emerging spectra (\emph{left column})
             and atmospheric structure (\emph{right column}) from the
             model parameters. The spectra are shown in energy flux
             times polar cap area, i.e., in luminosity units,
             for both polarization modes. Shown are from top to bottom
             how the spectral shape and the structure of the atmosphere
             depend on the mass accretion rate, $\dot{M}$ (a), the
             polar cap radius, $r_0$ (b), the cyclotron energy, $E_\cyc$
             (c), and the fraction of the accretion energy going
             into non-thermal collisional excitations, $f_\cyc$. The
             same baseline model is shown in purple in all panels. See
             text for a further discussion.
             }
      \label{fig:specs}
\end{figure*}

We now turn to studying the effects of the model parameters on the
photon spectra emerging from the neutron star atmosphere. We choose a
baseline model with parameters $\dot{M}=1.6\times10^{14}\gsec$,
$E_{\cyc}=70\kev$, $r_0=140\,\mathrm{m}$, and $f_{\cyc}=0$. The energy
flux times area is always shown for this model in
purple in Fig.~\ref{fig:specs}(a-d, left), with the corresponding
atmospheric profiles on the right. We then vary each parameter at a
time and show the spectra and the atmospheres relative to this model.
By decreasing the mass accretion rate compared to our baseline model
(Fig.~\ref{fig:specs}(a)), we obtain a lower bottom temperature and,
consequently, a weaker and softer thermal hump.
At a lower accretion rate, our iterative solution for the atmosphere
reaches the energy conservation (Eq.~\ref{eq:encons}) at a higher
$y_0$. Consequently, the volume in which most of the accretion energy
is released increases. This leads to a lower temperature also
in the entire upper part of the atmosphere, and, accordingly, to a
less active formation of the second hump, by reducing the Compton-$y$
parameter,
\begin{equation}\label{eq:compty}
    y_\mathrm{Compt}=\frac{2}{15}\frac{kT_{\mathrm{e}}}{m_{\mathrm{e}}c^2}\max{(\tau, \tau^2)},
\end{equation}
for a magnetized plasma. We note that Eq.~\ref{eq:compty} is applicable
only to the continuum formation and should be understood only in the
sense of dependence on electron temperature in case of partial energy
redistribution, which is responsible for the second hump formation.
A decrease in the polar radius (Fig.~\ref{fig:specs}(b)) leads to the
opposite effect on the atmosphere, as compared to a decrease in the
mass accretion rate since the volume of the entire energy release
decreases. Thus, the temperature rises throughout the atmosphere and
the thermal hump becomes much harder. However, as we show spectra in
energy flux times area units, it is clear that the luminosity is lower
for the model with a smaller polar radius. Both decrease in $r_0$ and
increase in $\dot{M}$, i.e, increase in flux density,
$\rho_0 v_\mathrm{ff}=\dot{M}/A$, leading to the higher stopping
length.

Figure~\ref{fig:specs}(c) shows that reducing the
cyclotron energy, i.e, the surface magnetic field, to 50\kev does not
affect the thermal hump. Even though the temperature of the upper part
of the atmosphere decreases, the strength of the second hump is not
reduced, although it is shifted to lower energies. The formation of
this spectral component directly depends on the cyclotron energy and
the temperature of the upper atmosphere. As we reduced the cyclotron
energy much more than the upper temperature, we left room for more
efficient energy redistribution near the cyclotron resonance, resulting
in strong wings and a deep line. It is important to mention, however,
that the width of the line only appears to be bigger than it is at
higher cyclotron energy because of the logarithmic energy scale.

Finally, we allow for non-thermal collisional excitations during
the plasma braking in the atmosphere (Fig.~\ref{fig:specs}(d)).
We increase the parameter $f_\cyc$ to $0.4$, so now 40\% of the
accreted energy goes into collisional excitations of the ambient
electrons, whereas the rest contributes to heating. This change
does not affect the stopping length, $y_0$, as we just redistribute
energy sources. For the atmospheric profile, it mainly lowers
the bottom temperature, softening the thermal hump. Simultaneously,
we see a rise in the second hump, as the shoulders become stronger
and the cyclotron line more shallow. Thus, the main effect of
adding a source of non-thermal cyclotron photons is a change in
the ratio of the thermal and the high-energy humps while
simultaneously reducing the line depth.

\section{Application to data: fitting \gx with
  \polcap}\label{sec:gx304-1}

The spectral shapes discussed in Sect.~\ref{sec:results} are in
qualitative agreement with the spectral shapes seen from accreting
neutron stars in the low $\dot{M}$ limit. However, only fitting the
model to real data will give a clear idea of its applicability. To
that end, we apply the \polcap  model to analyze deep observations of
the Be X-ray binary \gx made with \nustar and \textit{Swift}/XRT.

\gx was discovered in October 1967 during a balloon flight
from Mildura, Australia \citep{lewin1968a, lewin1968b}, and later by
\citet{mcclintock1971} in 1970. Observations with the Small Astronomy
Satellite 3 (SAS 3) in February 1977 revealed regular X-ray pulsations
with a period of about 272\,s \citep{mcclintock1977}. \gx
consists of a highly magnetized neutron star and the main sequence
fast-rotating Be-star V850 Cen. The system is located at a distance
of about 2\,kpc \citep{parkes1980, rouco2018}. At the time
of discovery the source showed the typical behavior of accreting
neutron stars in Be-systems, with regular type~I outbursts at an
orbital period of 132.5\,d, which are associated with the periastron
passages of the neutron star through the excretion disk of the Be star
donor \citep{priedhorsky1983}. \gx showed two periods of
regular activity. The first one lasted from the moment of the
discovery until 1980, when its flux dropped to quiescence
\citep{pietsch1986}, followed by a $28\,\mathrm{yr}$ long ``off''
state.

Renewed activity of \gx was found with \textit{INTEGRAL} in 2008
\citep{manousakis2008}, and the source stayed active until mid-2013.
During this time, using data from the Rossi X-ray Timing Explorer
(\textit{RXTE}) and \textit{Suzaku}, \citet{yamamoto2011} discovered
a CRSF with a centroid energy of ${\sim}54\kev$. Later analysis showed
that the CRSF is positively correlated with source luminosity
\citep{klochkov2012, malacaria2015}. \citet{malacaria2015} found the
CRSF detectable at all rotation phase bins and obtained a maximum
centroid energy of 62\kev, corresponding to a magnetic field strength
of $6.9\times10^{12}\,\mathrm{G}$. \citet{jaisawal2016} discovered a
significant variation of the cyclotron line parameters with
rotational phase. 

\citet{malacaria2017} showed that irregular activity of \gx is closely
related to the circumstellar disk evolution. During the outbursts,
when \gx had a typical luminosity of
$10^{36}\mbox{--}10^{37}\,\mathrm{erg}\,\mathrm{s}^{-1}$, the spectra are
well described by a cutoff power law \citep{yamamoto2011,
  jaisawal2016}. \citet{kuehnel2017} showed that intervals of
increased absorption column during this time are compatible with an
inclined Be disk, consistent also with the model for changes in the
geometry of the large-scale accretion flow \citep{postnov2015b}.
\citet{rothschild2017} suggested that spectral shape and positive
correlation of the CRSF with luminosity can be explained assuming the
accretion flow is in the collisionless shock braking regime. In the
quiescent state, however, a recent \nustar observation shows the
source to have a bimodal spectral shape with humps around 5\kev and
40\kev and comparable energy fluxes \citep{tsygankov2019a}, resembling
the spectral shapes discussed in Sect.~\ref{sec:results}. This makes
\gx an ideal candidate to test the \polcap model on observational
data.

We used a low luminosity \nustar observation of \gx, taken on
2018 June 3 (ObsID 90401326002) which is complemented by a
contemporaneous \textit{Swift}/XRT observation (ObsID 00088780001).
We processed the \nustar data with \texttt{NUSTARDAS} pipeline
version 2.0.0 with CalDB version 20201101 under \texttt{HEASoft}
V6.28. Source and background regions for both \nustar focal plane
modules (FPMs) are circles of 30\asec radius, centered on the source
and outside the point spread function wings, respectively. We reprocessed
the \textit{Swift}/XRT data with \texttt{xrtpipeline} version 0.13.5.
Source and background spectra were extracted using \texttt{XSELECT}
version 2.4 for a circular source region of 20\asec radius and an
annulus background region, with radii of 80\asec and 120\asec. We
use ISIS~v.~1.6.2 for our spectral analysis \citep{houck:00a}. We
applied optimal binning following the approach of \citet{kaastra2016}
and used Cash statistics \citep{cash1979} for spectral fitting. As
the observation was performed during a deep quiescent state, we
find the high energy band to be strongly background dominated above
${\sim}35$\kev. Therefore, we restrict the energy range used for
spectral fitting to 1--6\kev and 4--35\kev for
\textit{Swift}/XRT and \nustar, respectively.

To compare the \polcap model with the data we use the results from
the parametric study described in Sect.~\ref{sec:results} to generate
an additive table model that is compatible with \texttt{XSPEC} and
\texttt{ISIS}. Thus, the table model contains 144 spectra, between
which interpolation is possible. To obtain the flux on the detector,
we multiply the flux in the neutron star rest frame by the factor
$2\pi r_0^2/4\pi D^2$, for a distance $D=2\,\mathrm{kpc}$. We note
that this approach implies symmetric accretion onto two poles and
isotropic emission. Models in which these very simplifying assumptions
are relaxed will be presented in subsequent work. We fix the
gravitational redshift to $z=0.24$, as is appropriate to the
surface of a neutron star with $M_\ns=1.4\msun$ and
$R=12\,\mathrm{km}$. Finally, we account for photoelectric
absorption in the interstellar medium using the \texttt{tbnew}
model with the cross-sections by \citet{verner1996} and abundances
by \citet{wilms2000}.

The complete model used in our spectral fits is, in
\texttt{XSPEC}-like notation,
\begin{equation}
S(E) = \mathtt{detconst}\times\mathtt{tbnew}\times\mathtt{polcap},
\end{equation}
where \texttt{detconst} accounts for differences in flux calibration
between the \fpma and \fpmb detectors onboard \nustar, as well as
the difference between \fpma and \textit{Swift}/XRT calibration.
\begin{table}
    \caption{Parameters of the best-fit models of the pulse-averaged
    spectra shown in Fig.~\ref{fig:fit}.}
    \label{tab:fit_params}
    \centering
    \begin{tabular}{ll}
        \hline\hline
        Parameter & best-fit \\
        \hline
        $N_\mathrm{H}\,[10^{22}~\mathrm{cm^{-2}}]$ & $1.1$ (fixed) \rule{0pt}{2.5ex}  \\
        $\log_{10}(\dot{M}/[\mathrm{g}\,\mathrm{s}^{-1}])$ & $13.809^{+0.019}_{-0.023}$ \rule{0pt}{1.5ex}\\
        $r_{0}$ [m] & $\left(1.30^{+0.08}_{-0.07}\right)\times10^{2}$ \rule{0pt}{1.5ex}\\
        $E_\cyc$ [keV] & $57.6^{+1.1}_{-1.3}$ \rule{0pt}{1.5ex}\\
        $f_\cyc$ & $0.234^{+0.029}_{-0.041}$ \rule{0pt}{1.5ex}\\
        $z$ & $0.24$ (fixed) \rule{0pt}{1.5ex}\\
        $C_{\mathrm{FPMA}}$\tablefootmark{a} & $1$ (fixed) \rule{0pt}{1.5ex}\\
        $C_{\mathrm{FPMB}}$\tablefootmark{a} & $1.00\pm0.04$ \rule{0pt}{1.5ex}\\
        $C_{\mathrm{XRT}}$\tablefootmark{a} & $0.91^{+0.13}_{-0.11}$ \rule[-1.5ex]{0pt}{0pt}\\
        \hline
        Cash/d.o.f.& $276.03/221=1.25$\rule{0pt}{2ex}\\
        \hline
    \end{tabular}
    \tablefoot{The indicated uncertainties are at $1.6\sigma$ (90\%) confidence level.
              \tablefoottext{a}{Detector cross-calibration constants
              with respect to FPMA.}
              }
\end{table}
\begin{figure}
    \resizebox{\hsize}{!}{\includegraphics{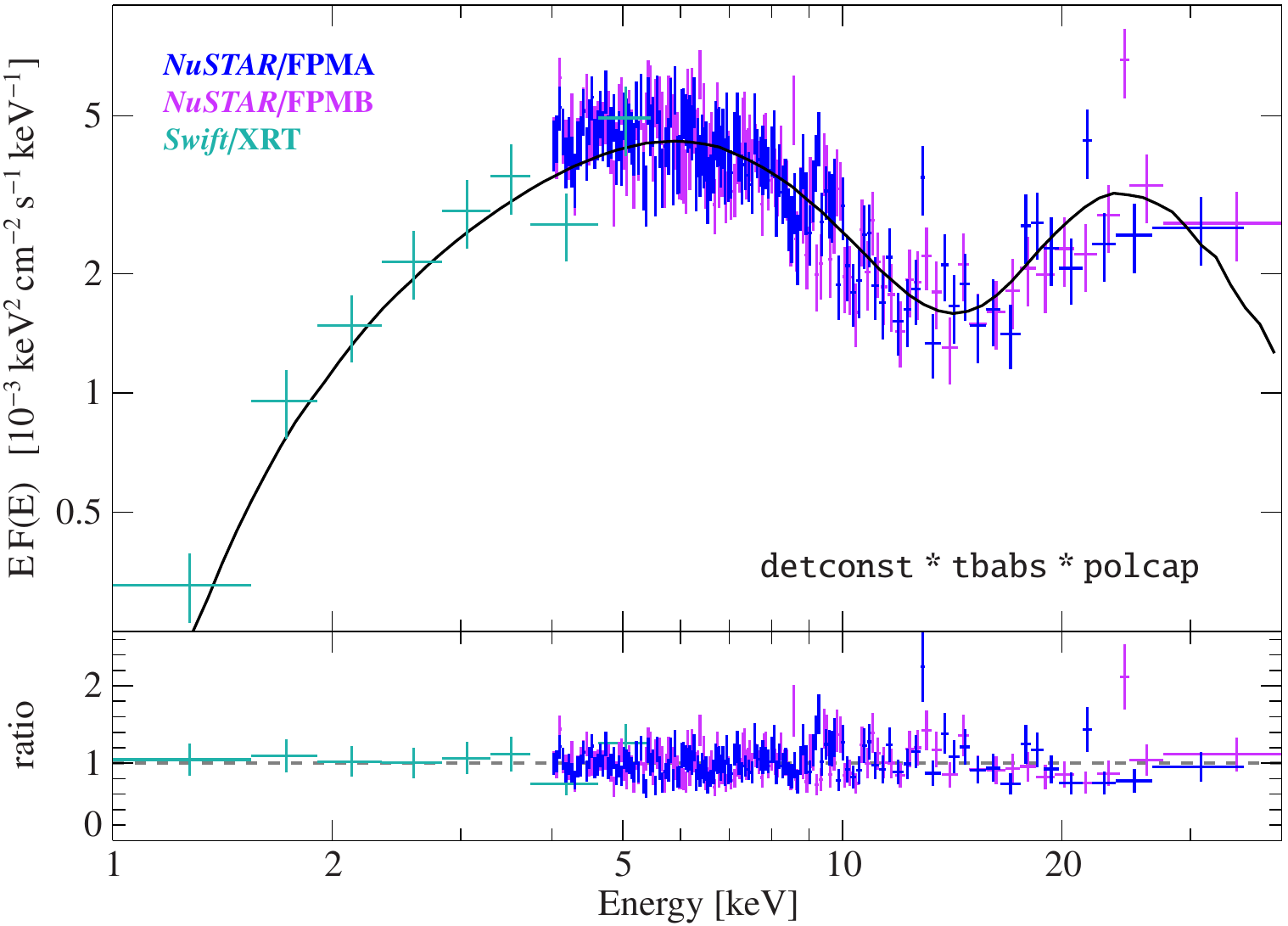}}
    \caption{Unfolded spectrum of the low luminosity observation of
             \gx in $EF(E)$ with the best-fit model (solid black
             line) for \nustar \fpma (blue), \fpmb (violet),
             and \textit{Swift}/XRT (cyan).
             For clarity we rebinned the spectra with larger
             bins for plotting than the ones used for the fit
              (\emph{top}).
             Ratio for the best-fit model (\emph{bottom}).
            }
    \label{fig:fit}
\end{figure}
We note that we explicitly do not re-scale our model during the fit
in order to ensure that we reproduce not only the spectral shape but
also the absolute value of the energy flux. Figure~\ref{fig:fit} shows
the unfolded phase-averaged spectrum and the best-fit model, the
parameters and statistics for which are presented in
Table~\ref{tab:fit_params}. Figure~\ref{fig:maps} provides confidence
plots to map the parameter space. The maps are calculated by stepping
through the two-dimensional grid corresponding to the two
parameters of the model and fitting the data at each point of the
grid. The obtained statistic is then compared to the best-fit one
(Table~\ref{tab:fit_params}). Contours for all parameters show the 
absence of degeneracies.

Overall, the spectrum shows the double-hump shape discussed
above. The high-energy excess in the spectrum is described by the
left-hand part of the ``second hump'', e.g., the red wing of the
cyclotron line formed in the hot non-homogeneous atmosphere due to
resonance Compton scattering.

The mass accretion rate from the best-fit model is
${\sim}6.4\times10^{13}\gsec$ and cyclotron energy is ${\sim}58\kev$,
which corresponds to a surface magnetic field
$B_0=5\times10^{12}\,\mathrm{G}$. The fundamental cyclotron line, in
this case, would be observed at ${\sim}58\kev/(1+z)\approx47\kev$.
This energy agrees with previous results, although in this case we
rather expect to see higher cyclotron line energy, breaking the
positive correlation of the CRSF with luminosity discussed by
\citet[][and references therein]{rothschild2017}, as emission comes
directly from the surface of the neutron star.
However, the cyclotron line energy range itself is not available for
the analysis in this observation as the data above ${\sim}35\kev$ are
dominated by background. The $E_\cyc$ parameter in this case is mainly
affected by the position of the left-hand part of the second hump,
which is not well constrained. In addition, the line position depends
on the angle under which the emission region is observed. Thus, by
using the spectrum averaged over the angles and avoiding the
relativistic ray tracing from the neutron star surface, we lose some
information about the line centroid energy. We will analyze it
in detail in our future paper on pulse-resolved extension of the
\polcap model (Sokolova-Lapa et al., in prep., paper~II hereafter).
The polar cap radius of ${\sim}130\,\mathrm{m}$ is close to the
estimate for a hot spot area for a dipole field \citep{lamb1973},
\begin{equation}
r_0\lesssim R_{\ns}\sqrt{R_{\ns}/R_{\mathrm{A}}}\approx160\,\mathrm{m},
\end{equation}
where $R_{\mathrm{A}}=\xi\mu^{4/7}(2\dot{M})^{-2/7}(2GM_\ns)^{-1/7}$
is the Alfv\'en radius, with dipole moment $\mu=B_0R_\ns^3$ and
$\xi=1$. In reality, the coefficient $\xi$ is close to unity, but
its exact value depends on a screening factor originating from
currents induced on a surface of an accretion disk and on magnetic
pitch at the inner edge of the disk \citep{wang1996}. The $f_\cyc$
parameter from the best-fit has a value of ${\sim}0.23$, i.e.,
$23\%$ of the energy of the accretion flow goes into non-thermal
excitations of Landau levels. As we noted before, this parameter
mainly affects the relative strength of the two spectral components
and can be influenced by the fact that the high-energy component
is not well constrained.
The total luminosity, under the assumptions stated above, is
$L=1.2\times10^{34}\,\mathrm{erg}\,\mathrm{s}^{-1}$.
\begin{figure*}
    \includegraphics[width=17cm]{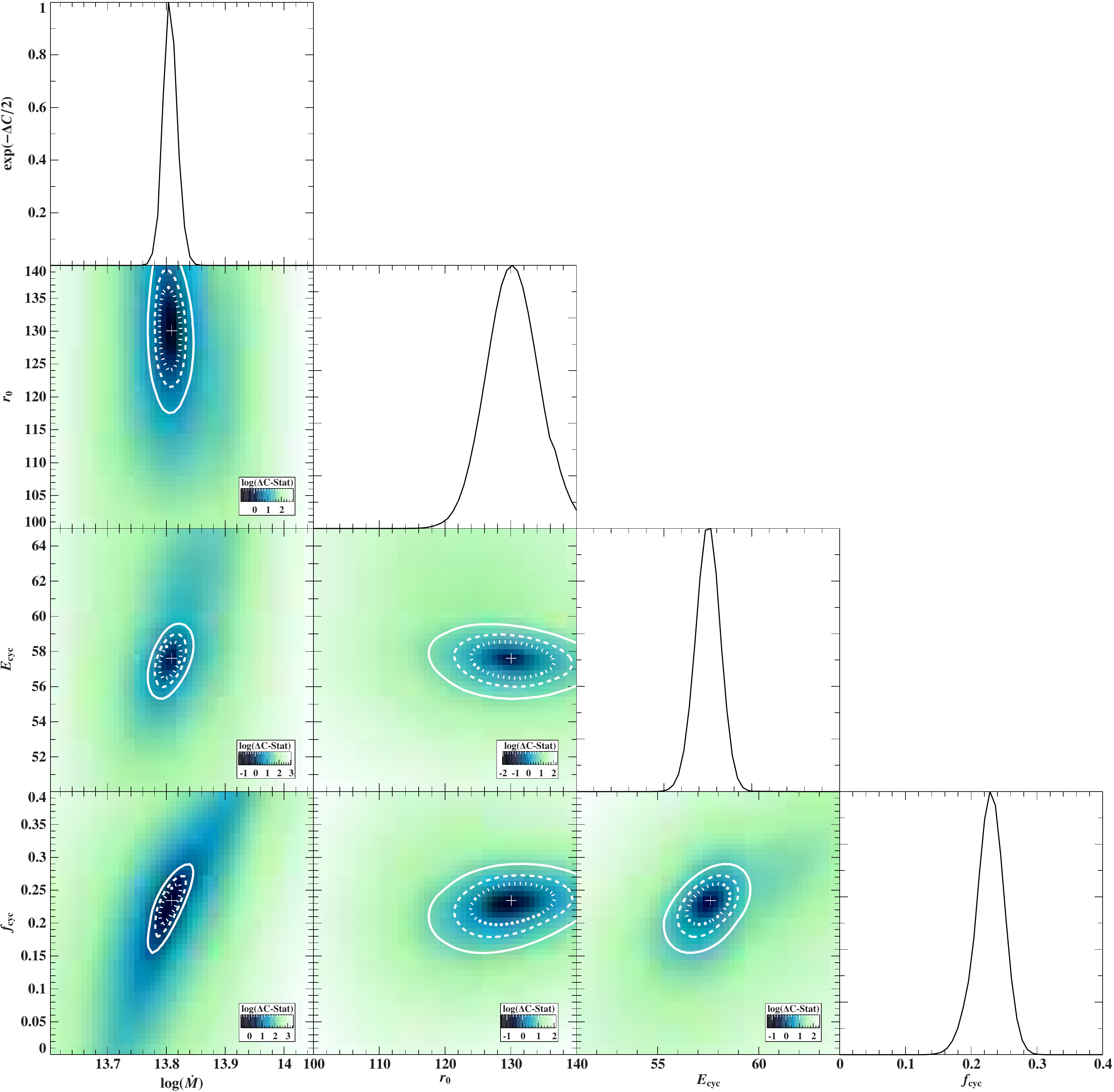}
    \caption{Parameter probabilities for the \polcap model applied to
      the low luminosity observation of GX\,304$-$1. Correlations for
      different parameters are shown in color as two-dimensional
      confidence maps, where dotted, dashed and solid contours
      correspond to the 99\%, 90\%, and 68\% confidence regions,
      respectively. Here ``$\Delta C$-Stat'' represents the statistic
      change compared to the best-fit. Side panels on the right show
      the one-dimensional parameter probabilities.}
    \label{fig:maps}
\end{figure*}

\section{Summary, Discussion, and Conclusions}\label{sec:disc}
\subsection{Summary}

In this paper we discussed our model of accretion at low $\dot{M}$
onto magnetized neutron stars. Despite the obvious simplifications in
our approach, we managed to obtain a structure of the atmosphere and
the radiation field that is internally consistent. This allows us to
understand, although partly in a parametric way, the physics of
accretion at low $\dot{M}$ and how these parameters influence the
emitted spectrum and its polarization.

We first showed that the atmosphere has a two-zone structure consisting
of a very hot layer which is mainly cooled by Compton scattering, and a
colder interior. Using angle-dependent, polarized radiative transfer
computations, we showed that this structure leads to a spectral energy
distribution that is characterized by two humps. The most important
results of our detailed radiative transfer are the following:
\begin{itemize}
\item The physical origin of the two-humped X-ray spectral shape is
  due to the strong polarization and temperature dependency of the
  Compton scattering cross section.

\item The lower-energy, thermal hump is mainly formed by the
  extraordinary photons that originate in the deep layers of the
  atmosphere.

\item The harder, second hump is due to resonant magnetic
  Comptonization in an atmospheric layer with a strong temperature
  gradient, and is modified by the cyclotron line in the upmost hot
  layer.

\item A fraction of the accretion energy is used to excite atmospheric
  electrons into higher Landau levels, which affects the relative
  luminosity of the two spectral components.
\end{itemize}
The formation of a high-energy excess by Comptonization in the
overheated part of the atmosphere with optical depth around unity is
well known for non-magnetic plasmas in neutron star low-mass X-ray
binaries \citep[see, e.g.,][]{deufel2001, suleimanov2018} and in X-ray
binaries with black holes \citep[e.g.,][and references
therein]{sunyaev1980,hua:95a,dove:97a,wilms2006}. For high
magnetic field values (with corresponding $E_\cyc\gtrsim40\kev$) this
region is also modified by the resonant nature of the
scattering. We expect this process to always be present in highly
magnetized atmospheres with the strong temperature gradient as formed
in the case of Coulomb collisions. As we showed in
Sect.~\ref{sec:results}, even without non-thermal collisional
excitations in the atmosphere (whose contribution is highly
conditional) the high-energy component in the spectra is produced
by resonant magnetic Comptonization.

Our model can quantitatively explain the observational data of \gx at
low luminosities. The parameters for the atmosphere that are obtained
with the model appear physically reasonable, with a polar cap radius
of ${\sim}130$\,m, mass accretion rate of $6.4\times10^{13}\gsec$,
and cyclotron energy of 58\kev at the surface of the neutron star.
The model requires about 23\% of the accreted energy to go into
non-thermal excitations of electrons into upper Landau levels.

These results open up various avenues for
further refining the study of the physics of accretion onto neutron
stars at low mass accretion rates. In the following sections, we
describe these in more detail, starting with the general atmospheric
structure in Sect.~\ref{dis:atm}, and then discussing in
Sect.~\ref{dis:colex} the expectations on the
collisional excitation of electrons as part of the process of plasma
stopping in the neutron star atmosphere. In Sect.~\ref{dis:colsh},
we address the alternative possibility of matter stopping at low mass
accretion rates, the collisionless shock. Finally, we summarize the
prospects for further development of the model in Sect.~\ref{dis:outl}.

\subsection{Atmospheric structure and proton stopping}\label{dis:atm}

As discussed above, the atmospheric structure of our model is typical
for collisionally heated atmospheres, with a hot optically thin layer
near the surface and a colder deep interior. In Sect.~\ref{sec:pardep}
we found that to a high degree the temperature of the overheated layer
depends on the proton stopping length.
In our model, $y_0$ is obtained indirectly, as a result of the
iterative procedure described in Sect.~\ref{ssec:calc}.
This approach, while greatly simplified, nevertheless allows for
some consistency between the atmosphere and the radiation field. It
also rules out the unphysical solutions and parameter values that
give too high or, conversely, unrealistically low emitting power of
the atmosphere.

The real picture of plasma braking in the neutron star atmosphere
is much more complicated, although it can be parameterized well in our
approach. It can be divided into three main sub-problems: (i) the
evolution of the distribution function of the falling protons after
entering the atmosphere, (ii) the balance between heating and cooling
in the atmospheric plasma and, finally, (iii) the radiation propagation
in the atmosphere. These problems, which are often complex by
themselves, in principle would need to be addressed in a
self-consistent manner. Advanced models \citep[e.g.,][]{meszaros1983,
  harding1984, miller1987, miller1989, nelson1993, nelson1995} couple,
to some degree, the proton stopping, energy balance, and radiative
processes, but still contain a number of serious simplifications,
especially regarding spectrum formation.

The Coulomb braking regime assumes that the incident flow gradually
loses energy inside the nearly static atmosphere of the neutron star
by proton-electron Coulomb collisions (with and without electron
excitation to a higher Landau level). The stopping power of the plasma
is also affected by the generation of the collective plasma
oscillations. The energy loss of protons falling along the $z$-axis,
resulting from both individual encounters and collective effects, is
\citep[see, e.g.,][]{nelson1993}
\begin{equation}\label{eq:enloss}
     \frac{\diff E}{\diff z}=-\frac{4\pi n_\mathrm{e} e^4}{m_\mathrm{e} v^2}\left(\Lambda_\mathrm{ind}+\Lambda_\mathrm{coll} \right) = -\frac{4\pi n_\mathrm{e} e^4}{m_\mathrm{e} v^2}\left(\frac{m_\mathrm{e} v^2}{\hbar\omega_\mathrm{p}}\right),
\end{equation}
where we implicitly define the Coulomb logarithms and where the term
$\Lambda_\mathrm{ind}$ includes both types of collisions, with and
without excitations. Equation~\ref{eq:enloss} also provides the
heating rate throughout the atmosphere, which is to be used in the
energy balance.

For a strong magnetic field, when only excitations up to 1--2 Landau
levels are possible, the stopping power of the atmosphere is smaller
than in the non-magnetic field case. \citet{meszaros1983} and
\citet{harding1984} found a stopping length of ${\sim} 20\gcm$ from
their stopping simulations in an inhomogeneous atmosphere for small
pitch angles collisions. \citet{miller1989} included Coulomb
collisions for arbitrary pitch angles and momentum transfer, obtaining
a stopping length of ${\sim} 40\gcm$. In this case, bremsstrahlung
remains the main source of photons, although a fraction of the
accretion stream energy is converted to cyclotron radiation produced
by the decay of collisionally excited electrons. We will discuss the
effect of this process in detail in Sect.~\ref{dis:colex}. The values
of $y_0$ found by our iterative simulations (Sect.~\ref{ssec:calc})
are lower than those obtained in the previous works on magnetic
atmospheres. This behavior is a consequence of the assumed energy
release profile along the $z$-axis of the slab (see
Eq.~\ref{eq:enbalexp}). The more detailed simulations of the Coulomb
braking in the atmosphere show that although the energy loss rate can
stay nearly constant throughout a significant part of the atmosphere,
it is rather lower than given by our simplified expression.

We note that we use non-magnetic coefficients to calculate the energy
balance (see Eq.~\ref{eq:lambb},\ref{eq:lambc}) in the current version
of the model, and thus the magnetic field strength on the surface of
the neutron star can influence the structure of the atmosphere only
indirectly through its effect on $y_0$ search in the numerical
procedure. However, similar to the model used here, more complex
calculations \citep{miller1989,harding1984} also result in
temperature profiles with a thin hot Compton cooling layer on top
and a much cooler lower region, similar to the non-magnetic case. The
atmospheric profiles obtained with our model successfully mimic this
self-consistent calculation for magnetic atmospheres, which suggests
that our approach is adequate for studying the spectrum emerging from
such atmospheres.

\subsection{Collisional excitation}\label{dis:colex}

The contribution of non-thermal collisional excitations to the process
of plasma stopping in the atmosphere of a neutron star strongly
depends on the strength of the source magnetic field (the higher the
field, the fewer excitations are possible) and the velocity of the
accretion flow, and is difficult to estimate in detail. For this
reason, we parametrize the contribution of collisional excitations to
the stopping process by introducing the parameter $f_\cyc$. We then
exclude this fraction of energy from the heating of the atmosphere
(Eq.~\ref{eq:enbal}), and directly convert it into cyclotron photons
(accounting for the Doppler broadening, Eq.~\ref{eq:cycsource}). This
approach allows us to investigate the influence of non-thermal
collisional excitations on the formation of the spectrum. We found
$f_\cyc=0.23$ for the case of the low luminosity observation of \gx,
i.e., about 23\% of the accreted energy is going into excitations of
electrons to the higher Landau level. In this section, we will try to
assess how reasonable this number is.

If the plasma approaches the neutron star surface at free-fall
velocity, $v_\mathrm{ff}$, then the maximum Landau level to which
electrons can be excited in electron-proton Coulomb collisions is
\begin{equation}\label{eq:nmax}
  n_\mathrm{max} = \frac{m_\mathrm{e}v^2_\mathrm{ff}}{2E_\cyc}.
\end{equation}
For our modeling of the spectra of \gx, with assumed parameters
$M_\ns=1.4\msun$ and radius $R_\ns=12\,\mathrm{km}$, and using the
cyclotron energy at the surface of the neutron star, $E_\cyc=58\kev$,
we find $n_\mathrm{max}=1$.

In detailed simulations of the proton stopping in neutron star
atmospheres, \citet{miller1989} showed that for magnetized atmospheres
with $n_\mathrm{max}$ up to 2 and for $B=5\times10^{12}\,\mathrm{G}$
the fraction of the accretion energy going into collisional
excitations of Landau levels is about 10\% \citep[Table 1, last
column, in][]{miller1989}. This is similar to the model of
\citet{nelson1993, nelson1995}, which predicts a high-energy excess
due to cyclotron photons as a result of the collisional excitations in
the atmosphere for $n_\mathrm{max}\gg1$. For high $n_\mathrm{max}$
\citet{nelson1993,nelson1995} predict a significant contribution of
the collisional excitations to the stopping process. However, for low
$n_\mathrm{max}=1,2$ their results agree with those of
\citet{miller1989} \citep[see, e.g., Fig.~7 in][]{nelson1993}. The
best-fit result of our parametrized description of non-thermal
collisional excitation is a bit higher than given by these simulations.
It can be caused by the fact that the high-energy component is
not well constrained in low luminosity observation of \gx, as well
as by ignoring the energy losses in the atmosphere, e.g., by the
generation of the collective plasma oscillations, within the framework
of our model. The latter can lead to a decrease in the strength of the low-energy component and result in the lower value of $f_\cyc$ found
by the best-fit.

We conclude that for highly magnetized atmospheres and low accretion
rates Coulomb collisions without excitations remain the main mechanism
of plasma braking, such that ${\sim}80\%$ of the accreted energy
is used to heat the atmosphere.

Finally, we comment on the consequences of these
estimates for observations of other sources. In Sect.~\ref{sec:pardep}
we showed that while a high-energy excess in the spectra is
formed in the absence of the seed cyclotron photons, $f_\cyc=0$, by
magnetic Comptonization, the real data require a contribution of
photons from non-thermal collisional excitations that amounts to
$f_\cyc=0.23$ to explain the spectral energy distribution of \gx.
We suggest a similar mechanism, e.g., resonant magnetic Comptonization
in the non-homogeneous atmosphere with a contribution of the
non-thermal cyclotron emission (depending on the source magnetic
field), for the formation of the high-energy excess of other
highly magnetized sources at very low luminosities. Thus, for \aplus,
with a slightly lower magnetic field, $E_\cyc=48(1+z)\kev$, following
Sect.~\ref{dis:colex} we obtain $n_\mathrm{max}=1$--2, depending on
the neutron star mass and radius. For \xper, estimates of the magnetic
field value vary. The proposed value from accretion torque models
yields $B\sim10^{14}\,\mathrm{G}$ \citep{doroshenko2012}. Assuming
similar spectral formation for \xper as for \gx and \aplus, as
suggested by \citet{tsygankov2019a, tsygankov2019b}, we expect the
cyclotron line to be located above ${\sim}100\kev$ which implies a
magnetic field $B\gtrsim9\times10^{12}\,\mathrm{G}$
\citep{mushtukov2020}.
Similarly, the magnetic field for \gro is estimated to be
$B\gtrsim 6.5\times10^{12}\,\mathrm{G}$, with a cyclotron line observed
at 78\,keV \citep{kuehnel2017b,bellm2014,yamamoto2014}, and a recent
claim of a significantly higher energy of ${\sim}90\,\kev$ \citep{ge2020}.
In the case of such high magnetic fields, collisional excitations in
the atmosphere are hardly possible. We therefore conclude that models
that explain the high-energy component mainly by collisional excitations
followed by cyclotron photon emission \citep{tsygankov2019b,
  mushtukov2020} require an unreasonably high contribution of this
process for the Coulomb braking regime, which is not supported by the
detailed studies on collisional stopping in magnetized atmospheres.

\subsection{Collisionless shock}\label{dis:colsh}

So far we consider that Coulomb collisions in the neutron star's
nearly static atmosphere were responsible for the braking of the
downfalling plasma, as suggested by ZS69 for spherical accretion and
later developed by many others for non-magnetic \citep[see,
e.g.,][]{turolla1994, deufel2001} and magnetic \citep[][and references
therein]{kirk1982, miller1989} atmospheres. We note, however, that
\citet{shapiro1975} proposed an alternative scenario that also
develops at low accretion rates onto a magnetized neutron star. In
this scenario a standing adiabatic shock wave forms at some distance,
$h_\mathrm{s}$, above the surface due to rapidly grown plasma
instabilities \citep{bykov2004}. The post-shock velocity of the
accretion flow is greatly reduced, $v_\mathrm{s}\sim v_\mathrm{ff}/4$
\citep{shapiro1975}. The matter therefore loses a significant fraction
of the kinetic energy passing through the narrow transitional region.
This energy release results in hot electrons behind the shock that
cool down mainly by the Compton effect, similar to the upper layer of
the Coulomb heated atmosphere. \citet{bykov2004} simulated the origin
of collisionless shocks for intermediate mass accretion rates,
$\dot{M}\sim10^{15}\mbox{--}10^{16}\gsec$.

Further studies are required to make a clear distinction between the
two regimes, stopping in the atmosphere by Coulomb collisions and by
passing through a collisionless shock. The important test here is the
behavior of the cyclotron line with luminosity. For the case of plasma
braking in the neutron star atmosphere, the line stays constant with
changing mass accretion rate since the height of the atmosphere is
almost independent of the accretion rate. In contrast, the height of
the collisional shock decreases with increasing mass accretion rate,
which can result in a shift of the line forming region to higher
magnetic field values. \citet{rothschild2017} applied this model to
explain the positive correlation of the cyclotron line with luminosity
in the spectra of \gx at intermediate mass accretion rates. Similarly,
\citet{vybornov2017} studied the variability of the cyclotron line and
the continuum hardness ratio with luminosity in the spectrum of
Cepheus\,X$-$4 based on the collisionless shock model. However, they
assumed that the emission from the aftershock region has the typical
power law spectrum.
A more accurate study of the spectral formation under the assumption
of a collisionless shock is required. Another possibility
to distinguish between collisionless shock and Coulomb braking regimes
is studying the pulse profiles that are expected to be different for
the case of an emitting hot spot and an accretion column.

\subsection{Outlook}\label{dis:outl}

Despite the simplifications outlined in the previous
sections, our model of accretion onto magnetized neutron stars at low
$\dot{M}$ already gives a good description of the observations.
Further modeling work and analysis of observations, however, is needed
to test the model and its limitations. First of all, based on
our discussion of the contribution of the collisional excitations to
the spectral formation in Sect.~\ref{dis:colex}, we argue that
more careful simulations of the stopping process in the column are
needed. In addition, these need to be compared with data from other
highly magnetized sources at low luminous states, e.g., \aplus, \gro,
and \xper, in order to test the consistency of the contribution of
non-thermal collisional excitations with an estimate of the source
magnetic field.

A second extension of the current model would be to perform even more
careful simulations of the cyclotron processes. Specifically, the
effect of plasma polarization to the normal modes and the relativistic
effects on the scattering cross sections have to be included in the
radiative transfer.
\citet{alexander1989} showed that the relativistic cyclotron line is
sharper, narrower, and appears at slightly lower energies than the
non-relativistic one for the radiative transfer under the same
atmospheric conditions. We assume that this might influence not only
the line formation itself at the most superficial layers of the
atmosphere but also the near-resonant scattering that forms a high
energy continuum. In addition, the introduction of proper modeling of
stimulated scattering, rather than the approximate way used here,
could affect the high energy spectrum by slightly increasing the flux
of the second component. The consistency of the model can furthermore
be significantly improved by direct coupling of the radiative transfer
equation with the energy balance in the atmosphere, which we plan to
take into account in the future development of the model.

Finally, in this paper we used angular averaged spectra to study the
spectral formation and to analyze phase-averaged data of \gx. However,
our model already includes the information of angular distribution of
the emitted photons and, combined with the proper simulation of
light-bending near the surface of the neutron star, provides a tool to
study pulse profiles and phase-resolved spectra. In this way, the
future development of the \polcap model will allow for the study of
pulse profiles at low accretion rates, distinguishing the contribution
of both polarization modes. This will be the subject of paper~II in
this series.

\begin{acknowledgements}
  This research has been partially funded by DFG grant 1830Wi1860/11-1
  and RFBR grant 18-502-12025. This research has been supported by the
  Interdisciplinary Scientific and Educational School of Moscow
  University ``Fundamental and Applied Space Research''. Astrophysics
  research at NRL is supported by NASA. J.A.G.\ acknowledges support
  from NASA Astrophysics Theory Program grant 80NSSC20K0540, and from
  the Alexander von Humboldt Foundation. C.M.\ is supported by an
  appointment to the NASA Postdoctoral Program at the Marshall Space
  Flight Center, administered by Universities Space Research
  Association under contract with NASA. Part of this work is based on
  public data from the Swift data archive and has made use of data
  obtained with NuSTAR, a project led by the California Institute of
  Technology, managed by the Jet Propulsion Laboratory, and funded by
  NASA. We acknowledge the use of the NuSTAR Data Analysis Software
  (NuSTARDAS) jointly developed by the ASI Science Data Center (ASDC,
  Italy) and the California Institute of Technology (USA). We also
  acknowledge the use of the XRT Data Analysis Software (XRTDAS)
  developed under the responsibility of the ASI Science Data Center
  (ASDC), Italy. This research made use of NumPy
  \citep{oliphant2006guide}, SciPy \citep{2020SciPy-NMeth}, and
  Matplotlib \citep{Hunter:2007} libraries, as well as Astropy
  (\url{http://www.astropy.org}), a community-developed core Python
  package for Astronomy \citep{astropy:2013, astropy:2018}, and
  \texttt{ISIS} functions (\texttt{ISISscripts}) provided by
  ECAP/Remeis observatory and MIT
  (\url{http://www.sternwarte.uni-erlangen.de/isis/}).

\end{acknowledgements}

\bibliographystyle{aa}

\newpage
\begin{appendix}
\section{Code verification}\label{app:tests}

In this section, we verify the validity of our radiative transfer
code, \texttt{FINRAD}, by applying it to some problems with known
analytical or numerical solutions, for different types of atmospheres.

\subsection{Pure scattering atmosphere}

\begin{figure}
    \resizebox{\hsize}{!}{\includegraphics{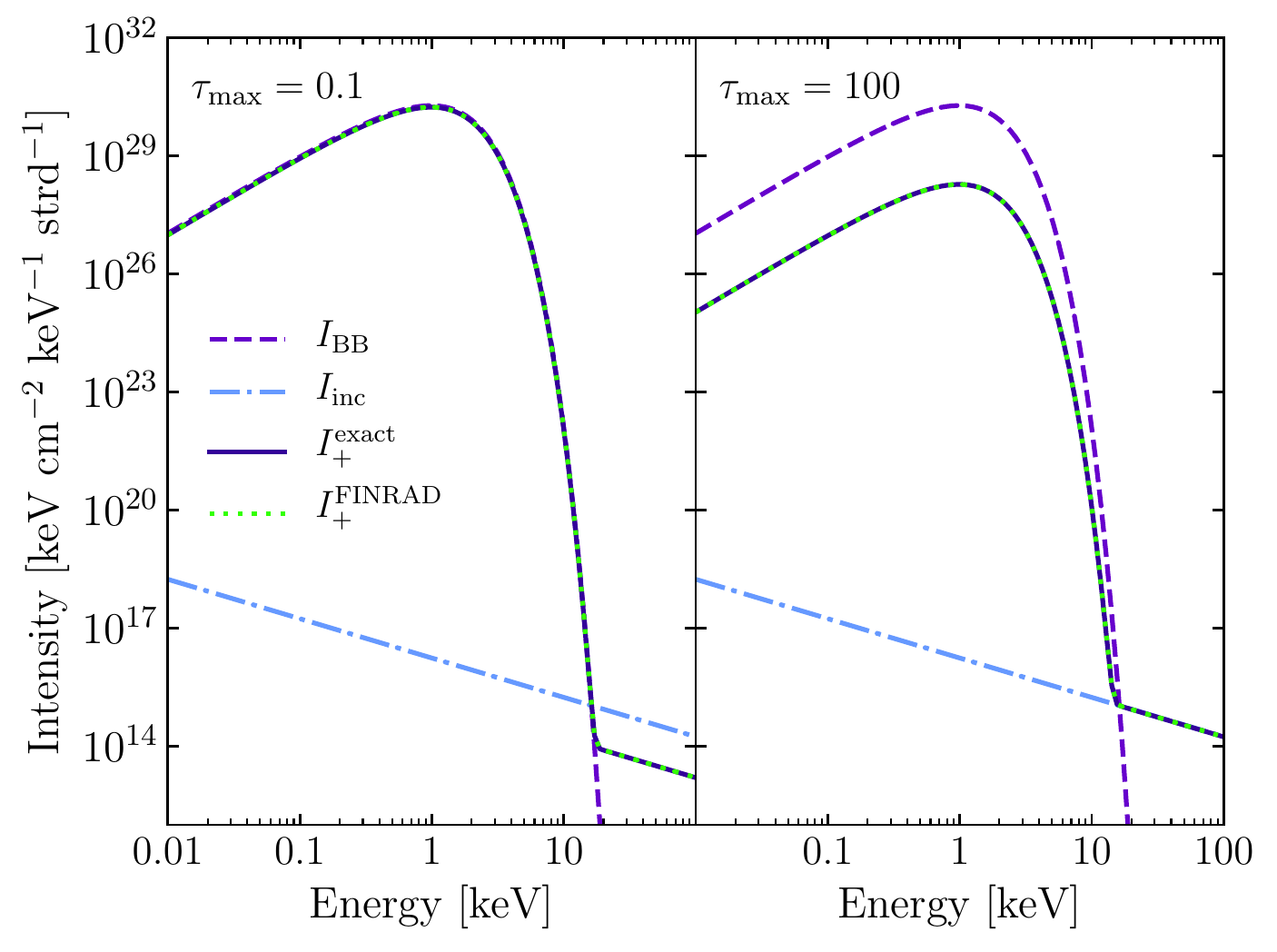}}
    \caption{Intensity of the radiation emerging from a
             homogeneous, slab-like atmosphere with pure isotropic
             scattering at an angle of $\theta=60^\circ$. The slab is
             illuminated by $I_\mathrm{inc}\propto E^{-1}$ from the top
             and by black body radiation from the bottom,
             $I_\mathrm{BB}(kT)$ with $kT=0.35$ keV.
             The results are presented for two different total optical
             depths: $\tau_\mathrm{max}=0.1$ (left) and
             $\tau_\mathrm{max}=100$ (right).
            }
    \label{fig:test1}
\end{figure}

We first consider the radiative transfer in a pure scattering
homogeneous atmosphere. The atmosphere is a slab, illuminated by
radiation with power law energy distribution, $I_\mathrm{inc}$, from
the top and by blackbody-like radiation from the bottom,
$I_\mathrm{BB}$. We consider isotropic Thomson scattering, therefore
the source function is $S_{E}=J_{E}=\int_0^1{u(E,\mu,\tau) d\mu}=u_E$.
In this case, the radiative transfer equation becomes
\begin{equation}
  \mu^2 \frac{\partial^2 u(E,\mu,\tau)}{\partial\tau^2} = 0
\end{equation}
with the straightforward solution
\begin{equation}
  u(E,\mu,\tau) = C_1 + C_2 \tau.
\end{equation}
where the constants can be found from the boundary conditions. At the
top of the slab ($\tau=0$) we set $I_{-} = I_\mathrm{inc}$, whereas at
the bottom ($\tau=\tau_{max}$) $I_{+} = I_{bb}$. The final solution is
therefore
\begin{equation}
  u(E,\mu,\tau) = \left( \frac{I_\mathrm{BB} - I_\mathrm{inc}}
  {\tau_{max} + 2\mu}\right)
  \left[ \tau - (\tau_\mathrm{max} + \mu)\right] + I_\mathrm{BB}.
\end{equation}
The intensity of the radiation emerging from the top of the slab can
be found by taking into account that
$u(E,\mu) = \tfrac{1}{2}(I_{+} + I_\mathrm{inc})$ (Eq.~\ref{eq:ufunc}).
Then
\begin{equation}\label{eq:test1}
  I^\mathrm{exact}_{+}(E,\mu) = I_\mathrm{inc}
  \left[ \frac{\tau_\mathrm{max}}{\tau_\mathrm{max}+2\mu}\right]
  + I_\mathrm{BB} \left[\frac{2\mu}{\tau_\mathrm{max}+2\mu} \right].
\end{equation}
We solve the same problem with \texttt{FINRAD} code for one angle of
the photon propagation, $\theta=60^\circ$. Figure~\ref{fig:test1}
shows that the numerical solution obtained with \texttt{FINRAD} fully
agrees with the analytical solution from Eq.~\ref{eq:test1}.

\subsection{Self-emitting slab with magnetic Compton scattering
and absorption}

\begin{figure}
    \resizebox{\hsize}{!}{\includegraphics{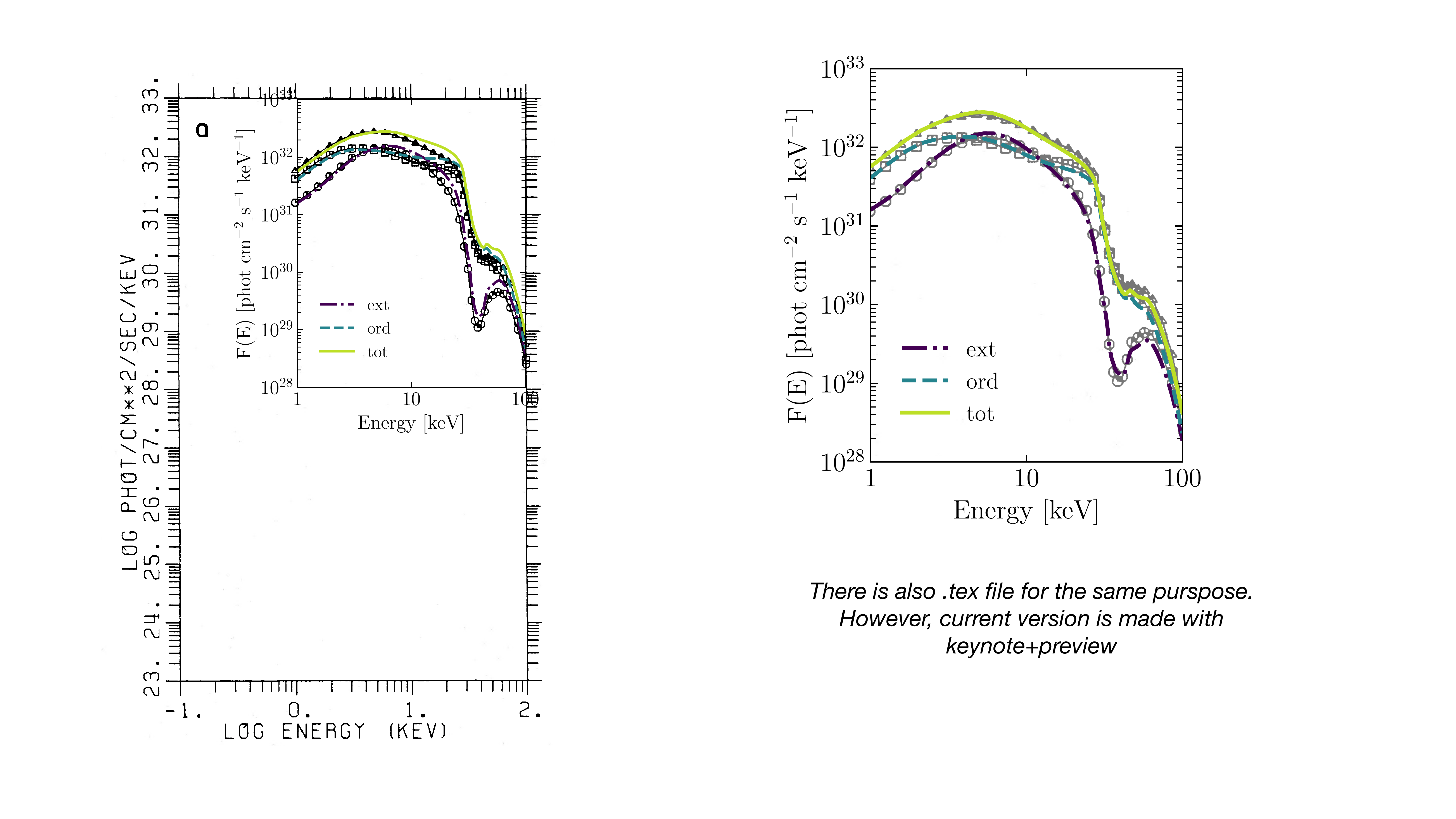}}
    \caption{Flux emerging from a homogeneous self-emitting atmosphere.
      Grey markers (circles, squared, and triangles) denoting
      extraordinary, ordinary, and the sum over two polarization
      modes, respectively, are from \citet[][Fig.~3a]{meszaros1985a}.
      Colored lines are the result of our simulations with the
      \texttt{FINRAD} code. }
    \label{fig:test2}
\end{figure}

To verify our code for the case of Compton scattering with partial
energy and angular redistribution in a magnetized atmosphere, we
reproduce calculations of \citet{meszaros1985a}. Specifically, we
consider a self-emitting homogeneous slab with scattering and
free-free absorption and emission. Cross sections of the processes are
given by the formulas in Sect.~\ref{ssec:sctabs}, but with mixed
plasma and vacuum normal modes of photons propagating inside the
medium. We solve Eq.~\ref{eq:rtrans1} in two polarization modes.
Following \citet{meszaros1985a}, the source function is given by
Eq.~\ref{eq:sourcef}, omitting, however, the additional source of the
cyclotron photons, $S_\cyc$. The only source of photons in the
atmosphere is free-free emission. The upper boundary condition is that
there is no illuminating radiation from above
\begin{equation}
  \mu \frac{\partial u}{\partial z} = -\frac{1}{2}I_{+}.
\end{equation}
There is no incident radiation from below as well. In this case, due
to the symmetry of the problem, only a half of the slab can be
considered, with a lower boundary condition set at the middle of the
slab
\begin{equation}
  \frac{\partial u}{\partial z} = 0.
\end{equation}
We follow Sect.~IVa of \citet{meszaros1985a}, assuming that the
magnetic field is perpendicular to the surface of the slab with
$E_\cyc=38\kev$. The slab of thickness $R=100\,\mathrm{cm}$ has a
homogeneous structure with density
$\rho=0.5\,\mathrm{g}\,\mathrm{cm}^{-3}$ and a temperature of
$kT=8\kev$. We simulate the radiative transfer with four angles
chosen, as before, according to the double Gauss-Legendre quadrature
($\theta=21^\circ$, $48^\circ$, $71^\circ$, and $86^\circ$). We show
the resulting angle-integrated flux in Fig.~\ref{fig:test2} together
with the solution obtained by \citet[][their Fig.~3a]{meszaros1985a}.
There are only very small deviations between both solutions, probably
due to the different energy grids chosen.

\end{appendix}

\end{document}